\documentclass[a4paper,onecolumn]{quantumarticle}
\pdfoutput=1
\usepackage[utf8]{inputenc}
\usepackage[english]{babel}
\usepackage[T1]{fontenc}
\usepackage{amsmath,amsfonts,amssymb,amsthm}
\usepackage{dsfont}
\usepackage{graphicx}
\usepackage{hyperref}
\usepackage{subcaption}
\usepackage{tikz}
\usepackage{cite}
\usepackage{footnote}
\usepackage{csquotes}

\begin{document}
\newcommand{\ket}[1]{| #1 \rangle}
\newcommand{\bra}[1]{\langle #1 |}
\newcommand{\scalar}[2]{\langle#1| #2 \rangle}
\newcommand{\modsq}[1]{|#1|^2}
\newcommand{\follows}{\, \Rightarrow \,}

\newcommand{\columnIV}[4]{\begin{pmatrix} #1 \\ #2 \\ #3 \\ #4 \\ \end{pmatrix}}

\newcommand{\columnV}[5]{\begin{pmatrix} #1 \\ #2 \\ #3 \\ #4 \\ #5 \\ \end{pmatrix}}

\newcommand{\graph}[1]{\textbf{#1}}

\newcommand{\complex}{\mathbb{C}}

\newcommand{\linefigure}[3]{}




\title{Enhanced quantum transport in chiral quantum walks}

\author{Emilio Annoni}
\affiliation{QTLab, Dipartimento di Fisica "Aldo Pontremoli", Università degli Studi di Milano, I-20133 Milano, Italy}
\email{emilio.annoni@studenti.unimi.it}
\author{Massimo Frigerio}
\email{massimo.frigerio@unimi.it}
\orcid{0000-0002-2821-0574}
\affiliation{QTLab, Dipartimento di Fisica "Aldo Pontremoli", Università degli Studi di Milano, I-20133 Milano, Italy}
\affiliation{INFN, Sezione di Milano, I-20133 Milano, Italy}
\author{Matteo G. A. Paris}
\email{matteo.paris@fisica.unimi.it}
\affiliation{QTLab, Dipartimento di Fisica "Aldo Pontremoli", Università degli Studi di Milano, I-20133 Milano, Italy}
\affiliation{INFN, Sezione di Milano, I-20133 Milano, Italy}
\orcid{0000-0001-7523-7289}

\maketitle 


\begin{abstract}
 Quantum transport across discrete structures is a relevant topic of solid state physics and quantum information science, which can be suitably studied in the context of continuous-time quantum walks. The addition of phases degrees of freedom, leading to chiral quantum walks, can also account for directional transport on graphs with loops. We discuss criteria for quantum transport and study the enhancement that can be achieved with chiral quantum walks on chain-like graphs, exploring different topologies for the chain units and optimizing over the phases. We select three candidate structures with optimal performances and we investigate their transport behaviour with Krylov reduction. While one of them can be reduced to a weighted line with minor couplings modulation, the other two are truly chiral quantum walks, with enhanced transport probability over long chain structures.

\end{abstract}


\tableofcontents

\section{Introduction}

\par{Quantum transport across discrete systems is a relevant topic of solid state physics and quantum information science \cite{mulken2011continuous,XU20086727,PhysRevE.79.011117,Tusun2019}, with several applications ranging from modern electronics to quantum algorithms 
and biology \cite{Santiago2020,Li2020,Kochaniak200917700}. The natural framework to discuss those phenomena is that of quantum walks, 
where notions from graph theory, Markov processes and quantum mechanics naturally fit together to provide a consistent framework \cite{Tumulka2006126,Nam2022}.}

\par{Originally, quantum walks were conceived as a direct quantum generalization of classical continuous time Markovian processes\cite{PhysRevA.48.1687}, and for this reason inherited from them a dynamics described by a real symmetric Hamiltonian.
A general and wide set of structures has been already analyzed and characterized with this approach \cite{mulken2011continuous}, while more recently the possibility of complex tunneling amplitudes has been considered \cite{cqw1,cqw2,frigerio2021generalized}. This sets the birth of a novel line of research, referred to as Chiral Quantum Walks, investigating the dynamics of discrete quantum systems governed by general, complex-valued Hamiltonians that reflect the topology of an underlying graph structure. The presence of complex amplitudes breaks the time reversal symmetry of real Hamiltonians and can often also break spatial symmetries in quantum transport tasks, whence the name \emph{chiral}; in addition, these evolutions acquire new degrees of freedom in the form of the arguments of those complex numbers. Tweaking their values strongly influences the transport properties of those systems, and may lead to tunable, directional and enhanced transport among the sites of the underlying graph \cite{cqw1,cqw2,frigerio2021generalized,PhysRevA.105.032425,10.1116/5.0146805,FRIGERIO202328}.}\\

\par{In the present paper, after an introduction to the theory of chiral quantum walks, we introduce the classes of structures we propose 
to enhance quantum transport. The sample of graphs tackled in this work consists in chain-like structures, namely linear graphs and their direct generalizations where each node may be substituted by a more complex unit. Using numerical simulation we characterize the behaviour 
of a simple but general set of units and observe how their properties generalize to longer chains thereof. Through this analysis, we uncover general trends regarding both peak transferred probability and transfer time in the bigger chain structures, providing also rigorous criteria to define our figures of merit; in particular, we see that optimality with respect to phase parameters and topology seems to be an inherited property of the units that generalizes to chains. After selecting a few optimal candidates among the examined chain-like graphs, we study their transport behaviour with Krylov reduction, and justify their performance with analytical methods involving their spectra, highlighting the role of chirality. We also unveil a specific chiral symmetry manifested by our optimal Hamiltonians and entailing the possibility of transporting quantum superpositions of localized states while preserving the relative phase.}

\par{All our findings point to the fact that quantum transport is best tackled in a coherent way, making full use of the multiple available paths, whose enhanced interference can be helped by chirality through a tight relation with the chosen topology.}


\section{Continuous Time Quantum Walks on graphs}
\label{sec:qw}

\begin{figure}[ht]
\centering
    \begin{subfigure}{0.26\textwidth}
        \includegraphics[width = \textwidth]{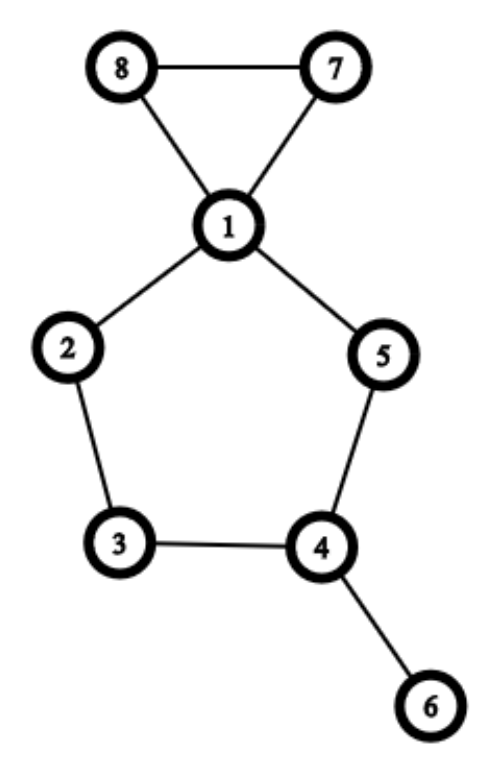}
        \caption{}
    \end{subfigure}
    \begin{subfigure}{0.5\textwidth}
    \renewcommand{\arraystretch}{1.6}
        \begin{equation*}
           \begin{bmatrix}
            4   &   -1  &   0   &   0   &   -1   &   0   &   -1   &   -1   \\
            -1   &   2  &   -1   &   0   &   0   &   0   &   0   &   0   \\
            0   &   -1   &   2   &   -1   &   0   &   0   &   0   &   0   \\
            0   &   0   &   -1   &   3   &   -1   &   -1   &   0   &   0   \\
            -1   &   0   &   0   &   -1   &   2   &   0   &   0   &   0   \\
            0   &   0   &   0   &   -1   &   0   &   1   &   0   &   0   \\
            -1   &   0   &   0   &   0   &   0   &   0   &   2   &   -1   \\
            -1   &   0   &   0   &   0   &   0   &   0   &   -1   &   2   \\
            \end{bmatrix}
        \end{equation*}
        \caption{}
        
    \end{subfigure}

    \caption{A numbered graph (\graph{a}) and its Laplacian matrix (\graph{b})}
    \label{fig:ex_graph}
\end{figure}

In order to formally define quantum evolutions on discrete structures, it is necessary to lay down the notion of a simple graph. A graph $\mathcal{G} = (V, E)$ is a pair of sets, where $V$ is the (finite) set of vertices and $E$ is the set of edges, whose elements are pairs of distinct elements of $V$. The cardinality of $V$ is the \emph{order} of the graph, and the number of edges involving a given vertex $v$ is the \emph{degree} of the vertex $v$ in $\mathcal{G}$. The graph is \emph{simple} if there is at most one edge connecting two generic vertices, and it is \emph{undirected} if the pair defining each edge is unordered. Let $\mathcal{G}$ be a simple graph of order $N$ and assume a given numbering of its vertices. Then one can construct an $N \times N$ symmetric matrix $\mathrm{A}$, called the \emph{adjacency matrix} of $\mathcal{G}$, whose entry $(j,k)$ is $1$ if vertices $j$ and $k$ are connected and all the others are $0$ (in particular $\mathrm{A}_{jj} = 0$). If $\mathrm{D}$ is the diagonal matrix whose entry $\mathrm{D}_{jj}$ is the degree of vertex $j$ for $j = 1,...,N$, the \emph{Laplacian matrix} of $\mathcal{G}$ is given by $\mathrm{L} = \mathrm{D} - \mathrm{A}$ (see Fig. \ref{fig:ex_graph}).

A continuous-time quantum walk on a graph is a discrete quantum system whose Hilbert space is generated by the states localized on the vertices of the graph and whose evolution is governed by an Hamiltonian embodying the connectivity of the graph, typically chosen to be its Laplacian or adjacency matrix.
To see how this goes, it is useful to first consider a classical random walk on a graph. Its classical state space corresponds to the simplex of probability vectors in $N$ dimensions, where $N$ is the number of vertices of the graph. The extremal states are those configurations in which the classical walker is localized in a precise vertex. Being $\underline{p}$ the vector of probabilities to be in each state, and $\mathrm{L}$ the Laplacian matrix of the system, the classical continuous time Markovian process is guided by the following equation:

\begin{equation}
\frac{d}{dt}\underline{p} = - \mathrm{L} \underline{p}
\end{equation}

The fact that $\mathrm{L}$ is bistochastic ensures that normalization and positivity of $\underline{p}$ are preserved. This evolution equation can be directly constrasted with Schrödinger's equation\footnote{For the sake of computational and exposition simplicity in this thesis we will absorb both $\hbar$ and the hypothetical energy proportionality constant of the Hamiltonian in the time scale, therefore both $H$ and the coordinate $t$ will always be regarded as adimensional. This way our analysis is not related to any particular energy scale, but one can easily restore the dimensionality of time as soon as one is faced with a specific system with defined energy levels.}:

\begin{equation}
\frac{d}{dt}{\ket{\psi}} = -i  \mathrm{H} {\ket{\psi}}
\end{equation}
where, in the site basis, the quantum state becomes a vector of amplitudes $\underline{\psi} \in \mathds{C}^{N}$. Apart from dimensional factors, here the edge between the two matrices is explicit. The standard association \cite{mulken2011continuous,portugal2013quantum,wong2016laplacian,kendon2006quantum}  amounts to choosing either $\mathrm{H} = \mathrm{L}$ or $\mathrm{H} = \mathrm{A}$. Notice that for regular graphs, i.e. graphs whose vertices all have the same number of neighbours, these choices are equivalent because the Hamiltonians differ by a multiple of the identity matrix and a sign in front of $\mathrm{A}$, which can be shown to be immaterial for the quantum evolution. Both these choices entail a real Hamiltonian, a fact which has striking consequences on the dynamics. \\

\par{To set up the transport problem, one fixes a starting vertex $\ket{1}$ and evaluates the square modulus of its projection on a target site $\ket{f}$ after a Schr\"{o}dinger evolution generated by either $\mathrm{A}$ or $\mathrm{L}$. The derivative of this localization probability is also readily computed in Schr\"{o}dinger picture as below.}
\begin{equation}
 p_{1 \rightarrow f}(t) = \modsq{ \bra{1} e^{-i \mathrm{H} t} \ket{f} }
\label{for:ptrans} 
\end {equation}

\begin{equation}
\frac{ d}{dt} p_{1 \rightarrow f}(t) = 2\, \mathrm{Re} \left( \bra{1} e^{-i \mathrm{H} t} \ket{f} \bra{f} i \mathrm{H}e^{i \mathrm{H} t}\ket{1}\right)
\label{for:ptrans_deriv} 
\end {equation}

Maxima analysis  involves finding their location in time as zeroes of (\ref{for:ptrans_deriv}) and their subsequent evaluation with  (\ref{for:ptrans}). Zeroes of (\ref{for:ptrans_deriv}) have no analytical expression for most Hamiltonians, therefore the only available tool in the general case is numerical analysis. \\

Whenever $\mathrm{H}$ is a real matrix in the site basis, the transport probabilities between generic sites $j$ and $k$ have the following nontrivial symmetries:
\begin{align}
\label{eq:timereversal}
& p_{j \to k} (t) \ = \ p_{j \to k} ( - t) \\
\label{eq:spatialsymm}
& p_{j \to k} (t) \ = \ p_{k \to j} (t) 
\end{align}
Notice that they are equivalent, since $p_{j \to k} (t) = P_{k \to j} (-t)$ is always true for generic transition amplitudes arising from unitary dynamics with a time-independent Hermitian generator. Eq.(\ref{eq:timereversal}) means that transition amplitudes between sites generated by real Hamiltonians have a time-reversal symmetry, while Eq.(\ref{eq:spatialsymm}) implies that the transport {\em cannot be directional} with these generators: inverting the initial site and the target site leads to the same transport probability. \\

To overcome these limitations and to lift the unnatural constraint of real Hamiltonians, the more general construction of chiral quantum walk has been proposed \cite{cqw1,cqw2,frigerio2021generalized}. They are still generated by an Hermitian matrix, but keeping the condition that the square modulus of the off-diagonal entries is either $0$ or $1$, depending on the graph connectivity \footnote{It is of course possible to also introduce weights for the different transitions. However, this partially contrasts with the original idea of quantum walks, and it is a different type of generalization than the one we are considering in this work.}. A more extensive discussion on correspondence criteria can be found in \cite{frigerio2021generalized}. \\

The net result is that transition probabilities for chiral quantum walks are defined exactly as in Eq.(\ref{for:ptrans}), but now $\mathrm{H}$ is no longer a real matrix in the site basis, and it will be referred to as a \emph{chiral Hamiltonian} on the underlying graph $\mathcal{G}$. Its diagonal elements $\mathrm{H}_{jj}$ are unconstrained, real numbers with the obvious significance of an on-site potential landscape, while the off-diagonal elements still reflect the graph topology thanks to the constraint $\vert \mathrm{H}_{jk} \vert = \mathrm{A}_{jk}$, where $\mathrm{A}$ is the adjacency matrix of the graph. Therefore:

\begin{equation}
\mathrm{H}_{jk} = \mathrm{H}_{kj}^* = e^{i\phi_{jk}} \ , \ \ \ \ \ \ \phi_{jk} \in [ 0, 2 \pi )
\end{equation}
so that to each directed edge $(j,k)$ in $\mathcal{G}$ a complex phase $e^{i \phi_{jk} }$ is assigned, and reversing the orientation of the edge amounts to complex conjugation of the associated phase, in order to maintain the Hermitian constraint for $\mathrm{H}$. 

\par{Throughout this work, by \emph{applying a phase} we mean the process of substituting the entry of $\mathrm{H}$ relative to that edge with the chosen complex phase factor $e^{ i \phi}$. We emphasize here that the implementation of these systems has been studied in multiple contexts in the recent literature, e.g. under the name of \emph{synthetic gauge fields on lattices} \cite{boada2017quantum,PRA21Novo,aidelsburger2015artificial,dalibard2011colloquium,wer19,doi:10.1126/sciadv.aat3174,PhysRevApplied.16.054036}, both in cold atoms and in all-optical platforms, moreover for planar graphs (as all the ones considered in our study) they can also be directly realized, at least in principle, as tight-binding models for charged particles hopping between the sites of the graph in an underlying magnetic field.}

\subsection{Free phases}
\label{intro:free}

Distinct configurations of phases on the edges of a graph $\mathcal{G}$, corresponding to different chiral Hamiltonians, can lead to the same transport probabilities between every pair of sites. 
We can therefore factor the space of all possible phase choices, and therefore the set of all Hamiltonians with the same underlying discrete structure and the same diagonal entries, into equivalence classes such that the Hamiltonians inside each class share the same site-to-site transition probabilities. It can be shown \cite{cqw2} that two chiral Hamiltonians $\mathrm{H}$ and $\mathrm{H}'$ on the same graph $\mathcal{G}$ are equivalent in this sense if and only if they are related by a \emph{quasi-gauge transformation}\footnote{Since these transformations do not affect the diagonal elements, $\mathrm{H}$ and $\mathrm{H}'$ must already coincide on the diagonal. }:
\begin{equation}
\mathrm{H} \sim \mathrm{H}' \ \ \ \ \iff \ \ \ \mathrm{H}' \ = \ \mathbf{\Lambda}_{\underline{\alpha}} \mathrm{H} \mathbf{\Lambda}_{\underline{\alpha}}^{\dagger}
\label{eq:quasigauge}
\end{equation}
where $\mathbf{\Lambda}_{\underline{\alpha}}$ is a diagonal unitary matrix with $(\mathbf{\Lambda}_{\underline{\alpha}})_{jj} = e^{i \alpha_{j}}$ and $\underline{\alpha} \in [0, 2 \pi)^{N}$. It is a simple task to show that $P_{j \to k}(t) = \vert \langle k \vert e^{-i \mathrm{H} t  }\vert j \rangle \vert^{2} = {P'}_{j \to k}(t) = \vert \langle k \vert e^{-i \mathrm{H}' t  }\vert j \rangle \vert^{2}$ for $j,k = 1,..., N$. In this way, the phases of at most $N - 1$ edges can be cancelled; indeed, if we multiply $\mathbf{\Lambda}_{\underline{\alpha}}$ by an additional phase factor $e^{i \theta}$, the transformation in Eq.(\ref{eq:quasigauge}) applied to $\mathrm{H}$ will result in the same matrix $\mathrm{H}'$, therefore only $N-1$ among the parameters of $\mathbf{\Lambda}_{\underline{\alpha}}$ can matter. A tree graph with $N$ vertices has exactly $N-1$ edges, and it is indeed the case that all chiral Hamiltonians on it belong to the same equivalence class of the adjacency matrix (disregarding the diagonal elements). In particular, on tree graphs the properties in Eq.(\ref{eq:timereversal}) and Eq.(\ref{eq:spatialsymm}) hold also for chiral quantum walks. However, as soon as the number of edges is larger than $N-1$, there will be at least one loop in the graph, and there will be a phase that cannot be cancelled. It turns out that the equivalence classes can be characterized by the total phases along each loop of the graph, i.e. the product of the phase factors of all the edges involved in the loop, for a given orientation of it: two chiral Hamiltonians with the same diagonal entries are related by a quasi-gauge transformation if and only if all their loops total phases coincide. Consequently, loops phases are invariants for chiral Hamiltonians, and they will be the pivotal degrees of freedom that we will consider besides the graph topology when trying to optimize transport. \\

In general, if a graph $\mathcal{G}$ has $E$ edges, the actual number of relevant phase parameters for transport will be $E - N +1$. For planar graphs, the corresponding independent loops are easily identified, since they are related to the number of regions in which the plane is partitioned by the graph. Letting $F$ be this number of \emph{finite} regions, thanks to the Euler characteristic of planar graphs we have:

\begin{equation}
N-E+F = 1 \, \Rightarrow \, F = E- N + 1 
\end{equation}
therefore $F$ is indeed equal to the number of relevant phase parameters \cite{frigerio2021generalized,FRIGERIO202328}.

\section{Graphs notation}
\label{sec:graphnot}

\par
\begin{figure}[htp]
\centering

\centering
    \begin{subfigure}{0.33\textwidth}
        \includegraphics[width = \textwidth]{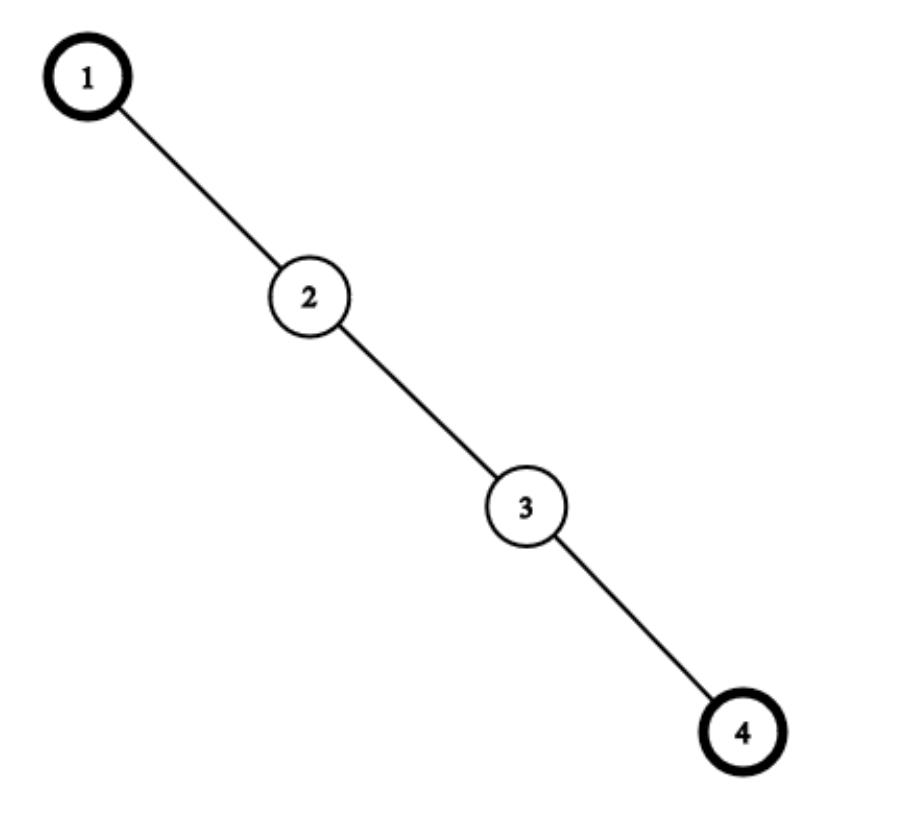}
    \end{subfigure}
    \begin{subfigure}{0.32\textwidth}
        \includegraphics[width = \textwidth]{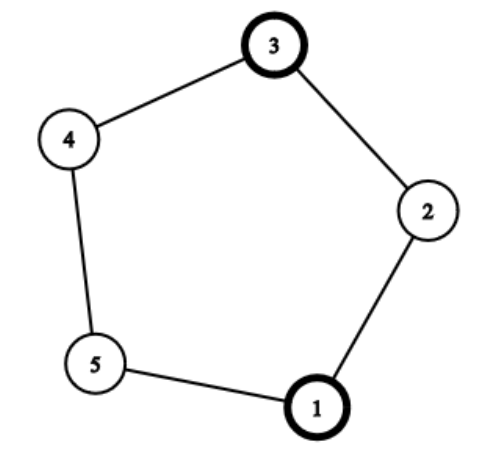}
    \end{subfigure}
    \begin{subfigure}{0.31\textwidth}
        \includegraphics[width = \textwidth]{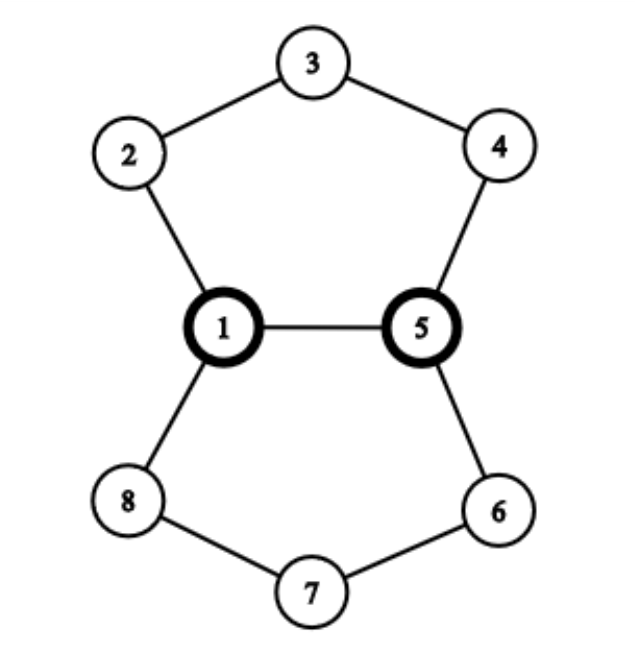}
    \end{subfigure}

\caption{
Three representatives of the three phase configurations addressed in this paper.
From left to right: \graph{P4}, \graph{C5}, \graph{DiC8(1-5)}. Start and end sites are highlighted.
}
\label{fig:1ph_graph}

\end{figure}

{Concerning the graph units that we shall use to construct chains, we will deal with three scenarios, showcased in Fig.\ref{fig:1ph_graph} and corresponding to $0$, $1$ and $2$ free phase parameters per unit, respectively. In this Section, we set our notation for these graph units, initial and target vertices for quantum transport and concatenation of units. A graph is always associated with a \emph{start site}, always addressed as $\ket{1}$, and a \emph{target site}, addressed as $\ket{f}$, which are respectively the state where the system is localized at $t = 0$ and the one in which we measure the (transferred) localization probability after the evolution.} \\

The simplest topologies have their own notation, which is formed from up to three letters that represent the name of the topology plus a number which determines the number of vertices. Example of the notation from simpler to more complex cases is shown in Fig.\ref{fig:chain_example} in Appendix \ref{app:chains}. Three kind of graphs are expressed in the following atomic notation:

\begin{itemize}

\item \underline{Path graphs}: (e.g. \graph{P9})
When the graph has no loops (i.e. it is a tree graph), all chiral Hamiltonians belong to the equivalence class of the adjacency matrix: no phase parameter have to be considered for transport analysis. This is the case of path graphs, denoted with \graph{P}, followed by an integer referring to the total number of sites.
Vertex numbering is progressive along the path and transport is studied between the first vertex and the last one. An eventual subscript (e.g. \graph{$P_{w}$}) indicates that the path has a speedup $w$, that is all the internal edges are set to have a weight $w$.

\item \underline{Cycle graphs}: (e.g. \graph{C4})

When a single loop is present in the unit, there is also a single phase parameter, which is entirely specified as the cumulative phase along the edges of the loop, once an orientation has been specified. The paramount example are cycle graphs, denoted with \graph{C} followed by an integer referring to the total number of sites. Vertex numbering is anti-clockwise and so is the convention for the sign of the only free phase, that is we say that the loop has a cumulative phase of $\phi$ if that value is obtained by traversing the edges following the path $1\rightarrow2\rightarrow\hdots \rightarrow N$.
Transport is always intended from the first vertex to the one that is diametrically opposite, the target vertex is thus well defined for even graphs( $1\rightarrow4$ in \graph{C6}) and is arbitrarily chosen to be the lower in number from the two opposites in odd graphs ($1\rightarrow3$ in \graph{C5}).
 
\item{\underline{Double cycle}: (e.g. \graph{DiC7(3,6)})

The third configuration considered will go under the name of \graph{DiC} because they are obtained by merging two cycles, followed by an integer, denoting the total number of sites, and two additional integers in brackets, denoting the pair of sites that are joined in order to split the larger cycle in two merged cycles. As a consequence, they result in two independent phase parameters. The example above refers to a 7-cycle graph (\graph{C7}) where vertices 3 and 6 are joined by an additional edge, effectively splitting the 7-cycle into a 5-cycle merged with a 4-cycle. When assigning phases to these graphs, we have two parameters to be considered. As for the cycles, we say the 5-cycle in \graph{DiC7(3,6)} comprising the vertices ${1,2,3,6,7}$ bears a phase $\phi_1$ if this is the value of the total phase when traversing the edges in increasing order of vertex number (as indicated in the list of vertices above). Similarly, the 4-cycle containing vertices ${3,4,5,6}$ bears a phase $\phi_2$ if this is the value of the sum of phases along oriented edges $3 \to 4$, $4 \to 5$, $5 \to 6$ and $6 \to 3$. We will use the notation $(\phi_1, \phi_2)$ to denote a phase configuration on a \graph{DiC} graph. Notice that with this convention the total phase on the external 7-cycle is exactly $\phi_1 + \phi_2$, while the phase on the added edge ($3 \to 6$ in our example) can always be put to $0$.  
}
\end{itemize}

The set of more complex graphs addressed used in this paper is built from the above list using three well defined operations, denoted as follows: 

\begin{itemize}

\item \graph{$h(\cdot)$} : (e.g. \graph{h(C6)} ) 

Represent "adding handles" to the graph, that is the operation of introducing two new vertices and joining the first one with the old \emph{start site} and the second one with the old target site. These new extended vertices take of course the role of the transport endpoints in the new graph. In the case of cycles, we attach the handles to maximally distant vertices on the cycle ($1$ and $4$ in \graph{C6} or $1$ and $5$ in \graph{C9}).

\item \graph{+} : (e.g \graph{C3+C4})

This operation between two graphs stands for joining the end site of the first unit to the \emph{start site} from the second unit by forming a new edge. The resulting graph has therefore no new vertices and counts one more edge. \emph{start site} in this new graph is the one that was \emph{start site} in the left operand while target site from the right operand becomes the new target site.

\item \graph{/} : (e.g \graph{C7/C5})

This is also an operation between two graphs where now the target site of the left operand and the \emph{start site} of the right operand are merged together. The resulting graph has the same number of edge of the union of the two old ones, but counts one less vertex. The new ends for transport are chosen exactly as in the \graph{+} operation.

\end{itemize}

Using this notation the graph structure in Fig \ref{fig:ex_graph}a may be denoted by \graph{C3/C5+P1}. Although not exhaustive, the overall
set of structures obtained with the above units and operations reflect the main sensible ideas of how one would assemble graphs for a transport purposes.
Another important feature is that the set of operation we chose to compose graphs is simple enough so that no new loop or fork can ever be created in the process of composition, meaning that overall transport still take place on a linear structure. 

\section{Criteria for quantum transport}
\label{ssec:1ph_t}

As discussed in the previous Sections, with chiral quantum walks the configuration of phases on a given graph come into play as 
a new degree of freedom  to influence the evolution. To determine the optimal configuration of phases, some criteria to define and 
describe optimal quantum transport must be outlined. Generally speaking, we aim at obtaining: 
\begin{itemize}
    \item a transport probability from the starting vertex to the final vertex which is as close to $1$ as possible. Typically, a lower bound should be introduced in order to refer to a maximum (a peak) in the transport probability as a proper transport event.
    \item The shortest possible time for the transport to occur.
\end{itemize}
\par{The second condition is crucial: in general, we are dealing with discrete quantum systems with incommensurate spectra where interference phenomena at large times quickly become difficult to predict. It is physically meaningless to look for anomalous transport events at such large times since, on most experimental platforms, coherence could be spoiled well before, and also because some kind of minimum speed constraint is usually implicit in transport tasks. We will see that a suitable choice of the phases may significantly improve the 
figures of merit characterizing quantum transport on a graph according to both of these goals. }\\

\begin{figure}[h!]
\centering
\includegraphics[width=0.95 \textwidth ]{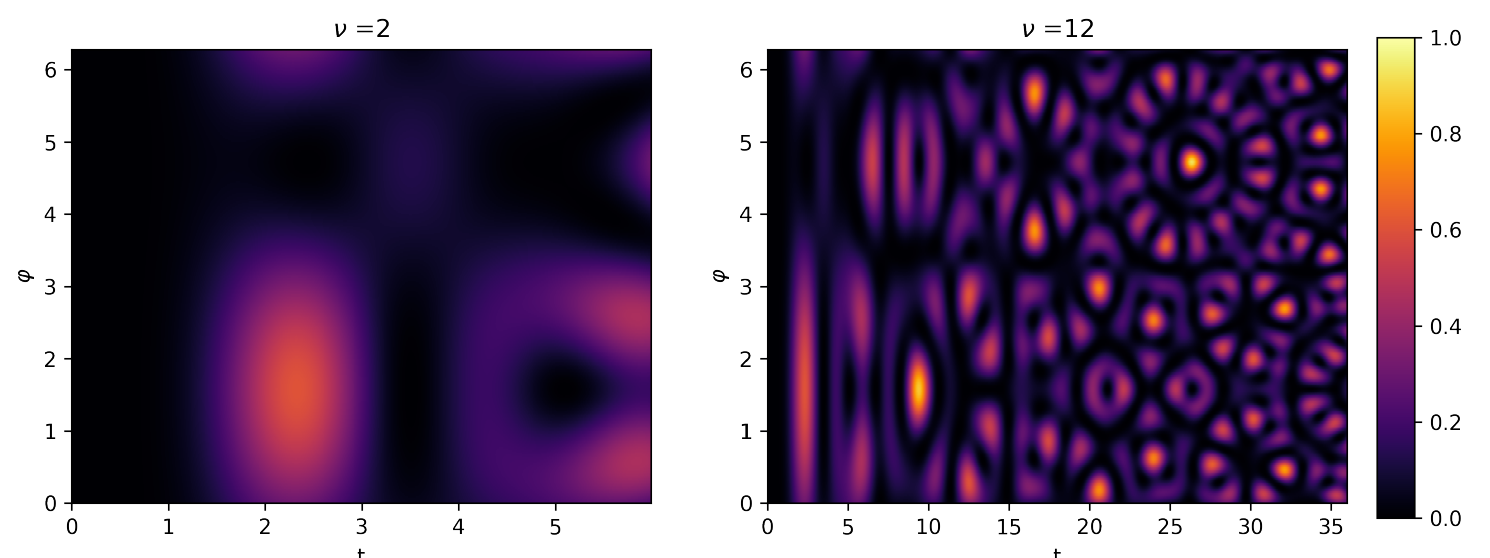}
\caption{
Evolution of localization probability  on target site for \graph{C7} as a function of time (x) and phase applied to the cycle (y). The two plot represent the same evolution on 2 different time window, $\nu$ (see below) is respectively set to 2 (\graph{a}) and 12 (\graph{b}). It is evident how difference in applied phase may result in two completely unrelated evolutions.}
\label{fig:c7_evo}
\end{figure}

{The need for a thorough discussion of transport criteria is well illustrated in Fig.\ref{fig:c7_evo}, where we plot the transport probability from site $1$ to site $4$ as a function of time and phase for the graph \graph{C7}. Upon defining the transport time as 
the time at which the highest peak of transport probability occurs in the given time window, it is apparent that the value of the 
transport time is strongly dependent upon the width of the time window itself. 
When a short time interval is considered, the transport time is mildly dependent upon the phase value (left panel of Fig.\ref{fig:c7_evo}), whereas for longer time windows later transport events are selected by numerical optimization and the dependence of the transport time with the phase becomes highly nontrivial (right panel of Fig.\ref{fig:c7_evo}).}



It can be seen that, considering a wide time interval, multiple transport maxima occur, which are localized both in phase and time. This implies that relatively small changes in the phase parameter can induce sudden jumps in the transport event selected by the numerical optimization. To cure this ambiguity and instability in the individuation of the transport events, search criteria guiding the optimization must be established. A first approach would consist in picking a specific time window for the walker's evolution and search for the absolute maximum of the localization probability both in time and phase. 
A fair comparison also requires to chose the time window according to the graph size, in order to take into account the different 
distance between the starting and target sites. Based on the ballistic spreading observed in the continuous-time quantum walk 
on a path graph, we select a time window given by:
\begin{equation}
\label{eq:tmax}
T_{max} = \nu \cdot d
\end{equation}
The distance $d$ is defined here as the minimum number of edges in a path connecting the endpoints of the transport task, while $\nu$
may be interpreted as the inverse of the transport velocity on the given graph.
Here we point out that, since a continuous-time quantum walk can be mapped to the one-particle subspace of a spin chain with local interactions, the Lieb-Robinson bound applies \cite{Lieb1972,Hast04,Bravyi06,Chandr10} and it is pointless to try to decrease $\nu$ arbitrarily \footnote{This is strictly true when no weights are introduced on the edges; the energy-time generalized uncertainty relations always imply the possibility of decreasing $\nu$ by increasing the energy scale proportionally}.

\begin{figure}[htp]

\centering

\includegraphics[width = \textwidth]{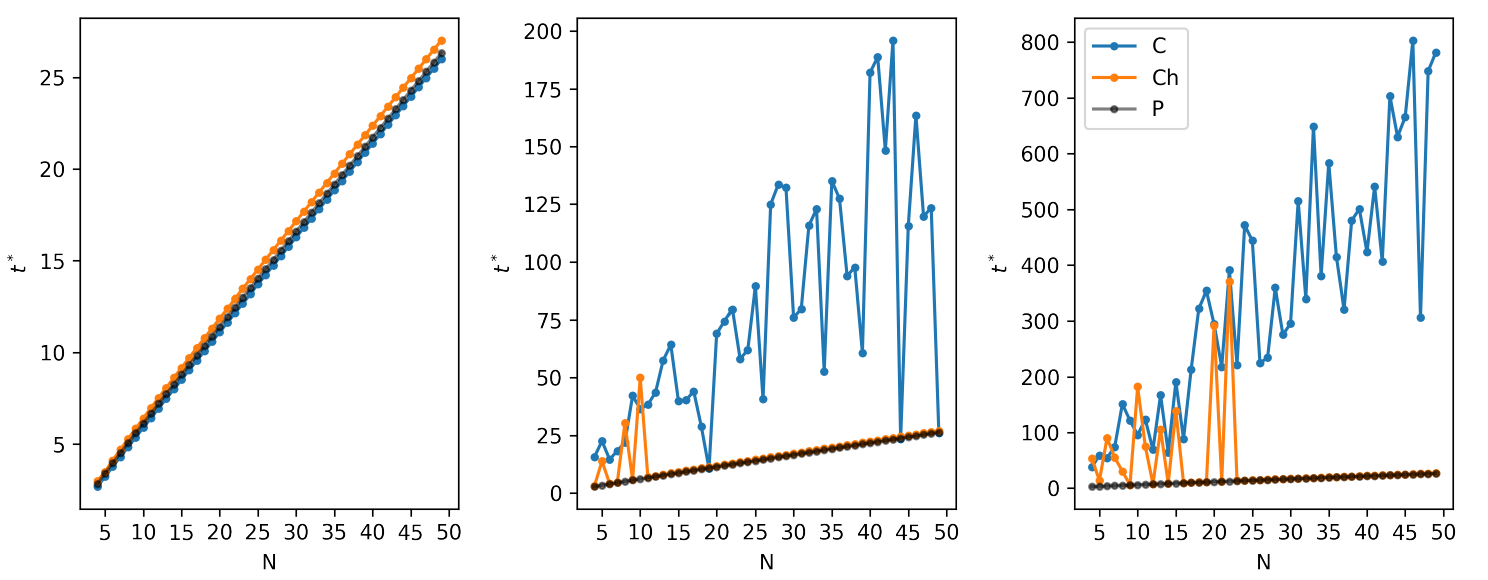}
\caption{
Transport time (corresponding to the first maximum of the transport probability) for a sample of graphs from the 
three classes \graph{P}(black), \graph{C}(blue) and \graph{h(C)}(orange), plotted as function of endpoint distance.
The search time-window for each graph scales linearly with distance. We set $\nu=1$ (a), $\nu=5$ (b), and $\nu=20$ (c). 
}
\label{fig:simple_t_prog}

\end{figure}

The ballistic behavior inspiring Eq.(\ref{eq:tmax}) is backed up by the results illustrated in 
Fig.\ref{fig:simple_t_prog}, where we show the transport time (corresponding to the first maximum of the transport probability) 
for a set of three simple topologies, namely \graph{P}, \graph{C}, and \graph{h(C)}.
In particular, when $\nu$ is $1$ the linear trend with distance is well established, while for 
$\nu = 10$ (right) the transport times oscillate wildly with distance, indicating that the 
numerical optimization is picking up long time features of the evolution where exceptionally 
high transport probability peaks can happen for certain values of $d$ in a highly unpredictable 
fashion. 

This approach also showcases interesting properties of \graph{C} and \graph{h(C)} families.
Odd and even cycles both follow a linear lower bound of transport times, however the difference in the symmetry of their endpoints apparently entails a constant shift in transport time, quantified later in Tab.\ref{tab:1ph}.
For this reason, to pick out the ballistic behaviour resulting from the choice of $\nu$, we choose to depict in Fig.\ref{fig:simple_t_prog} the \graph{C} and \graph{h(C)} graphs with an odd number of vertices only. \\

\par{As anticipated above, the transport times displaying a linear trend with the distance (for $\nu \simeq 1$) are those corresponding 
 to the first maxima of the transport probability. In light of this observation, a perhaps more consequential criterion to select transport 
events is to establish a threshold on probability, and pick the first local maximum as a function of time exceeding that threshold. 
In turn, this approach produces the same plot of Fig.\ref{fig:simple_t_prog}, where the linear scaling is most apparent. 
On top of that, the existence of a first wave of maxima with a distinct transmission velocity independent of the phase value, 
such as the one in Fig.\ref{fig:c7_evo} (left panel), is a general phenomenon of chiral quantum walks, as established by extensive 
numerical trials, also in more complex graphs. Overall, this criterion (a {\em transport event} occurs when the transport probability exceeds a given threshold for the first time) ensures a convenient trade-off between the two transport goals cited at the beginning of the Section.}

\par
\begin{figure}[htp]
\centering
\includegraphics[width =.48\textwidth]{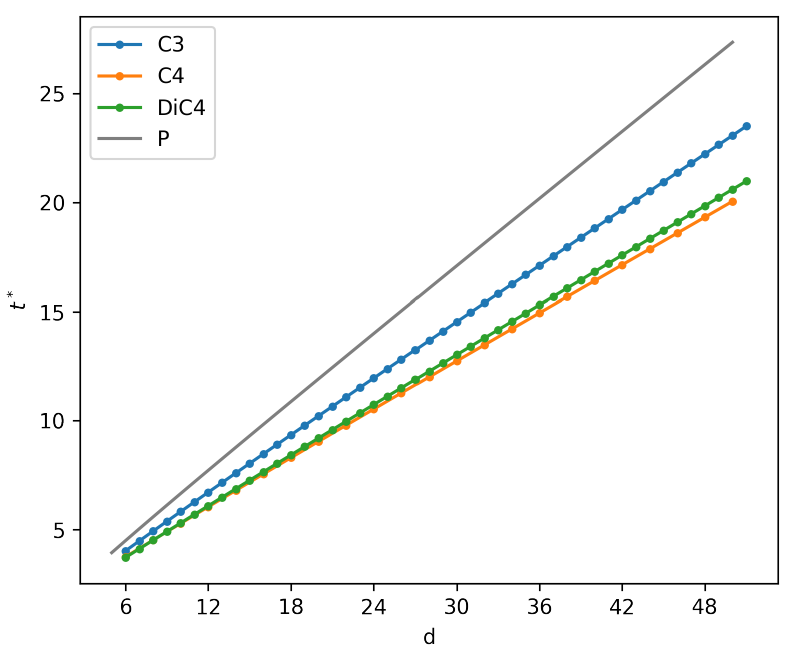}
\caption{
Transport time of first maxima (as obtained with the optimal phase value) for chains composed by 3 different units: \graph{C3}, \graph{C4}, \graph{DiC4(1,3)}. The black solid line represents the behaviour observed for \graph{P} (for reference). As it is apparent from the plot, all the conisdered chain topologies provide a speedup with respect to the reference. The horizontal axis displays the topological distance (directly related to the number of units), while the vertical axis represents the transport times of first maxima. 
}
\label{fig:chain_linear}

\end{figure}

{Additionally, when going from single units to chain graphs, the time-distance relationship for first maxima of transport still obeys a ballistic law, as shown in Fig.\ref{fig:chain_linear}. However, the transport time of the first maxima is influenced by the phase: the optimal value is the only one guaranteed to yield a clean linear behaviour and each case result in a different transmission speed.}

Our results also show that the optimal phase value for transport performance also results in the fastest transmission velocity, 
with no further compromise needed among our two transport goals. Transport time scaling for graph chains is noticeably improved
with respect to \graph{P} graphs: Krylov reduction (see section \ref{sec:extra_kry}) suggests that the multiple paths provided 
by the alternative topologies are the cause of the higher transmission velocity, a feature that has been observed for every unit 
in our sample. In the special instance of \graph{C4} chains without handles, one can prove that transport $\sqrt{2}$ time faster 
than \graph{P} graphs of the same length is achievable.
 
From Fig.\ref{fig:chain_linear}, it can be seen that the addition of handles, both here and in the \graph{h(C)} case, modifies 
the transmission velocity in a non trivial way since a \graph{C4} chain with handles does not show an exact $\sqrt{2}$ speedup, 
nor it seems to approach that value in the limit of a long chain. A time vs. distance linear regression has been carried out, and results are reported in Table \ref{tab:1ph} for single units of growing size (left table) 
and for chains made up of units (right table).  

\begin{center} 
\begin{table}[htp]
    \begin{subtable}{0.45\textwidth}
        \begin{tabular}{c||c|c|c|}
        Topology &	m &		q & $m/m_P$ \\
        \hline
        \graph{P}           &	0.522   &   1.454   &   1       \\
        \graph{C} (even)  &	0.532   &   0.541   &   1.02    \\
        \graph{C} (odd)   &	0.533   &	0.753   &   1.02    \\
        \graph{h(C)} (even) &	0.558   &   0.768   &   1.07    \\
        \graph{h(C)} (odd)  &	0.558   &   1.031   &   1.07     \\
        \end{tabular}
        
        \caption{}
        \label{tab:1ph}
    \end{subtable} 
    \begin{subtable}{0.45\textwidth}
        \begin{tabular}{c||c|c|c|}
        Unit &	m &	q   & $m/m_P$ \\
        \hline
        \graph{P}      &   0.522   &	1.454   &   1       \\
        \graph{C3}      &   0.451   &	1.310   &   0.86    \\
        \graph{C4}      &   0.378   &	1.488   &   0.72    \\
        \graph{C7}      &   0.538   &	1.246   &   1.03    \\
        \graph{C10}     &   0.471   &	1.508   &   0.90    \\
        \graph{DiC4}    &   0.395   &   1.350   &   0.76    \\
        \end{tabular}
        
        \caption{}
        \label{tab:chains}
    \end{subtable}
    
    \caption{Linear regression of the first maxima according to the formula $t=m\cdot d+q$, \graph{P} class values are used as a reference. The figures shown are related to the three 1- or-no-phase topologies (a) and to chains built from different units (b). All regressions 
    yield a coefficient of determination $R>0.999$}

\end{table}
\end{center}

These results validate the choice of the first maxima criterion to define the quantum transport event, thereby having to 
optimize only over the topology of the units and over the phase parameter(s).

\section{ Phase response}
\label{ssec:1ph_phase}

Once a criterion to individuate transport event is in place, the transport probability 
can be analyzed and optimized as a function of the free phases assigned to the loops of the graph.
\begin{figure}[h!]
\centering
\includegraphics[width = 0.88 \textwidth]{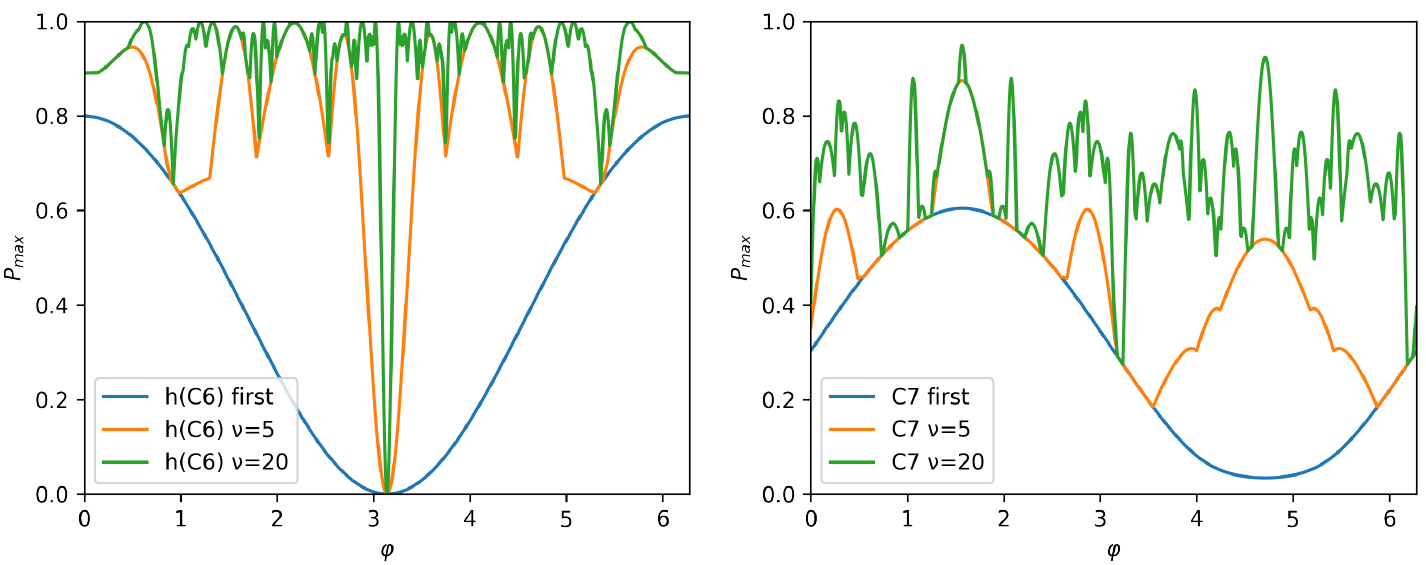}
\caption{
Plot of phase-dependent transport performance for \graph{h(C6)} (left) and \graph{C7} (right). For each 
graph we compare the result obtained using first maxima with what is obtained with maximization with bounds 
set up with different values of $\nu$. As for time performance the plot for each value of $\nu$ acts 
as a lower bound for any higher choice of it.}
\label{fig:perf_multiTC}
\end{figure}

In Fig.\ref{fig:perf_multiTC}, we show results for the graphs \graph{h(C6)} (left) and \graph{C7} (right), 
both with a single free phase parameter, which showcase the general trend for the whole 
\graph{C} and \graph{h(C)} families of graphs. As the time window for maximization gets wider, we observe 
an increase in the average performance for almost all phases, but also a progressively higher 
sensitivity to the specific value of the phase. This is yet another consequence of the unpredictable 
regime which governs the evolution of system with incommensurate spectra at large times. This 
also hints at the fact that any reliable implementation of quantum transport with chiral quantum walks should work in a suitable time regime that scales linearly with the length, so that the performance is not too fragile with respect to variations of the phase parameter.
Graphs with more elaborate structures made no exception to this behaviour, whereas the first maximum 
of transport probability show a smoother phase response throughout in all instances, with little to no 
effect on the best transport performance. For this reason, the subsequent analysis will be mostly focused 
on this latter class of events.

\begin{figure}[h!]
\centering
\includegraphics[width = 0.9 \textwidth]{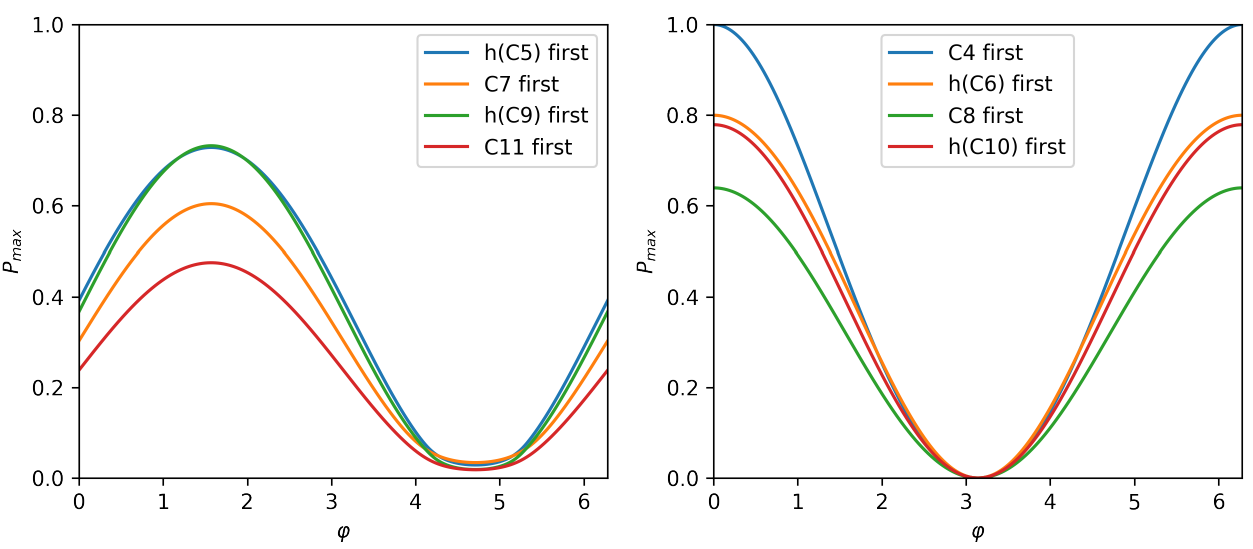}

\caption{
Phase dependent transport probability (first maxima) for \graph{C} and \graph{h(C)} graphs containing 
odd  (left) and even (right) cycles .
}
\label{fig:perf_even_odd}
\end{figure}

We can then characterize the phase response of \graph{C} and \graph{h(C)} classes as in Fig.\ref{fig:perf_even_odd}. As for the transport times, we observe two trends followed by graphs with odd or even number of vertices. In the even case, the optimal value of the phase is always $0$ (corresponding to a non-chiral quantum walk), whereas transport suppression at all times is obtained for $\phi = \pi$ \cite{Sett2019,cqw1,cqw2}. For odd cycles, the maximum is obtained for $ \phi =  \pi /2$, again independently of the specific size of the graph \footnote{We recall that for odd cycles with $2N+1$ vertices the target vertex is numbered $N+1$, while the starting vertex is numbered with $1$ as per usual. With $\graph{h(C)}$ the starting and target vertices are the endpoints of the handles, which are in turn attached to the cycle at a pair of maximally distant vertices.}

\par
\begin{figure}[h!]
\includegraphics[width = 0.9 \textwidth]{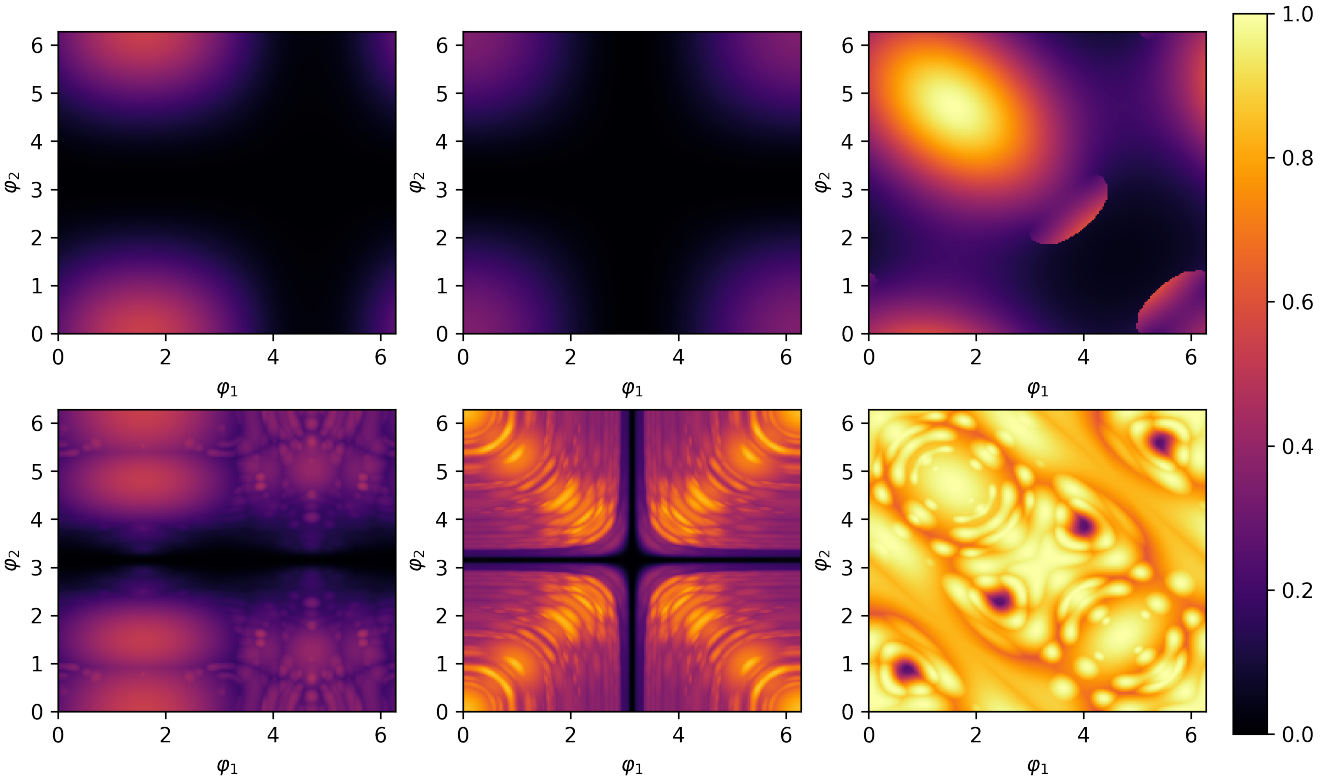}
\caption{
Transport probability as a function of the two free phases for \graph{C9/C4} (left), \graph{C8+C8} (center) and \graph{DiC4(right)} (c,f). The upper row show transport events corresponding to first maxima, whereas 
the lower row show results obtained by optimization over a linear time window with $\nu=10$.
}
\label{fig:2ph_vs}
\end{figure}

{Numerical analysis of graphs with two-loops indicate that the optimal phase values for 
each loop exactly combine to give the optimal two-phase configuration. An example can be seen 
in Fig.\ref{fig:2ph_vs}, where we show the transport probability as a function of the two phases 
for three chosen graphs, selecting just first maxima (upper row) or as a result of numerical 
maximization in a linear time window (lower row).
Search in a given time window returns a better (but less smooth) transport probability, and 
the optimal phase values for the single units (either $\pi/2$ or $\pi$ in Fig.\ref{fig:perf_multiTC}) 
result in local maxima for the two-loop graphs (Fig.\ref{fig:2ph_vs}).
First maxima events return an even clearer picture, where the globally optimal phases configuration 
corresponds exactly to the optimal choice of phase value for each cycle independently. The generality 
of this feature is backed-up by robust numerical evidence and holds for any tested combination 
of two-loop graphs.}

\par{A second relevant (locally optimal) configuration is encountered in more symmetrical situations 
when, for example, we bridge two identical cycles (see Appendix \ref{app:chains}).
First maxima transport probabilities obtained in such a scenario are shown in the last two columns of Fig.\ref{fig:2ph_vs}. There, the optimal phase values lie on the diagonal of the two-dimensional space of parameters 
due to the symmetry with respect to the swap of the two phases.
From this observation, one could make an educated guess that can guide our analysis in those cases where
more free parameters are present: the most relevant transport maxima are obtained for equal phases, 
especially if one focuses on first maxima.

Long chains made out of 1-loop units have been investigated in this way, and although numerical optimization 
over the whole space of phases parameters may further improve the transport 
performance, these observations make at least plausible, as we shall indeed show in the following, 
to treat the problem as a one-dimensional optimization problem over a single phase, equal for all 
the loops in the chain.} \\

\par{Concerning the ways to join two loops to form graphs with two free phases, we have examined 
three cases corresponding to: i) sharing a vertex, ii) bridging with an additional edge, and iii) 
merging along one common edge (see Appendix \ref{app:chains} for notation).
In this last instance, the resulting graph (\graph{DiC4(1,3)} for the case of only four vertices) allows for almost perfect transport at the first maximum as shown in the right column of Fig.\ref{fig:2ph_vs}. This is not particularly rather surprising, since it can be seen as a consequence of the distance between the starting vertex and 
the target one being reduced to $1$. Unlike the simpler \graph{P2} graph though, one would speculate that this kind of units could make better use of phase and exploit the interference of the three different paths also when used to construct chains. The structure \graph{DiC4(1,3)} is indeed the best performing chain unit according to both transport 
probability and transport time.
Also, the optimal phase configuration for this unit is $(\pi/2,-\pi/2)$, consistent with the result for single $\graph{C3}$ cycles.}

\par
\begin{figure}[h!]
\centering
    \begin{subfigure}{.38\textwidth}
        \includegraphics[width =\textwidth]{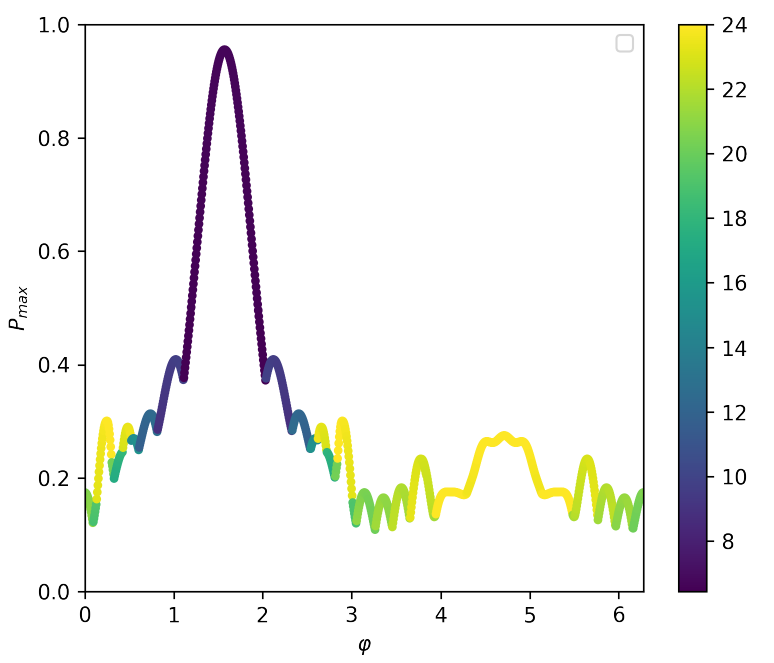}

    \end{subfigure}
    \begin{subfigure}{.37\textwidth}
        \includegraphics[width =\textwidth ]{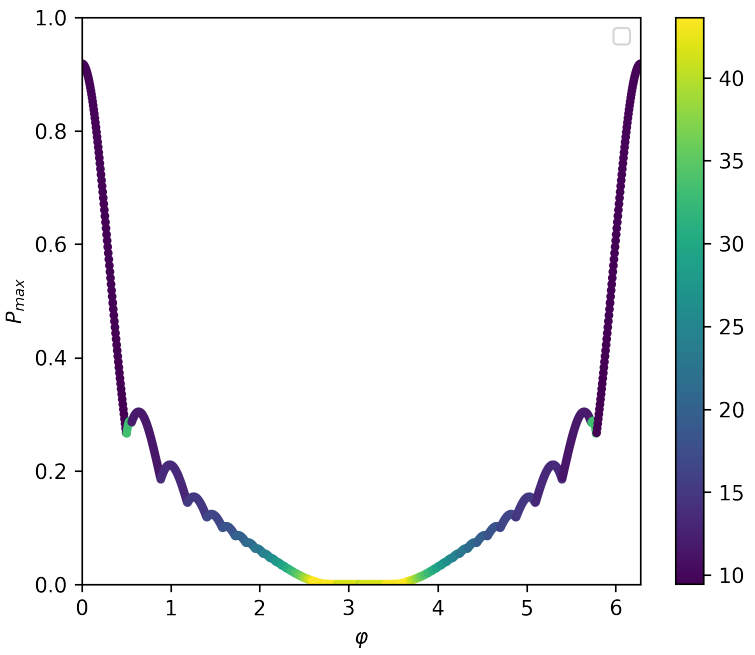}

    \end{subfigure}
\caption{
Transport probability as a function of the single phase parameter, for 10-units optimal chains (see App.\ref{app:chains}), with \graph{C3} units (left panel) and \graph{C4} units (right panel). Transport maxima are chosen with the linear time window criterion ($\nu$ is set to 2), transport time is drawn as color according to the shown colorbars.}
\label{fig:chain_perf}
\end{figure}

When dealing with general chain graphs, one has a large number of free phase parameters and the numerical optimization becomes quickly unstable due to this high dimensionality and to the wildly oscillating dependence of the transport probability with respect to all the parameters involved. Nevertheless, the results gathered above suggest that such a in-depth analysis is not necessary, as one can find very good transport events by setting the same phase configuration for each unit and then optimizing the transport probability at the first maximum over the few free parameters (one for each loop of the chosen unit). Results obtained in this way are shown in the two panels in Fig.\ref{fig:chain_perf}, where we plot the transport probability as a function of the phase for chains composed by \graph{C3} and \graph{C4} units, showcasing the general trends for chains of odd and even cycles, respectively.
The unequivocally optimal phase, corresponding to the peaks in the plot, indeed coincides with the optimal phase 
for the respective unit (0 for \graph{C4} and $\pi/2$ for \graph{C3}).
It is worth noticing that the transport times of the selected maxima, denoted by the color shading, highlight that the optimal phase value (with respect to transport probability) also results in the quickest transport.
An analogous result has been found using \graph{DiC4} as a unit: as shown in Fig.\ref{fig:2ph_vs} (right), the best configuration for the two phases of this unit is located on the anti-diagonal at $(\pi/2, -\pi/2)$, and still optimal 
when extended to a chain.

\section{ Asymptotics for chain performance}
\label{sec:hyer}
\begin{figure}[!h!]
\includegraphics[width =0.95\textwidth]{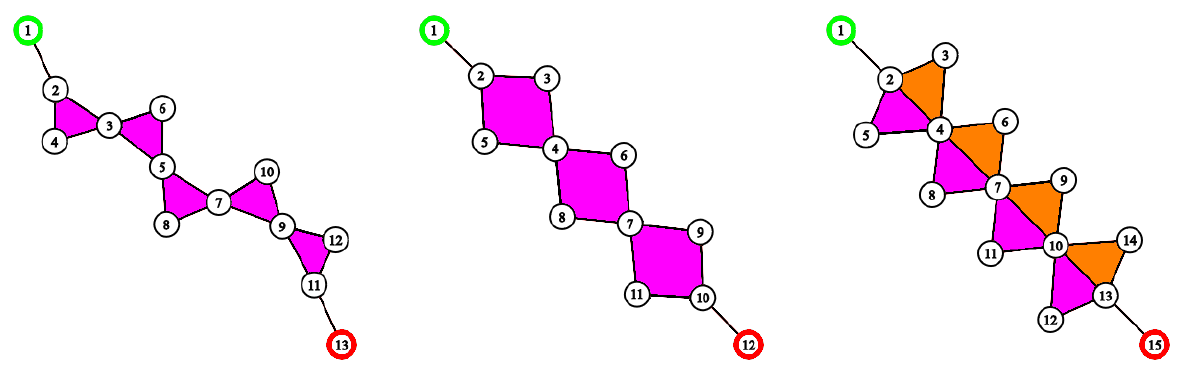}
\caption{
Representatives from each of the three best chain families found in this work. In order: \graph{C3} chain with 5 units, \graph{C4} chain with 3 units and \graph{DiC4} chain with 4 units. Start and target site have been highlighted, the independent phase loops of each unit have been colored
}
\label{fig:best_chain}
\end{figure}

Our analysis so far suggests that chains of small 1-loop or 2-loop graphs joined one to the next by a 
shared vertex perform well as candidates to achieve quantum transport over long distance.
Moreover, the optimality of the homogeneous phase configuration for chains, combined with the observations on computational feasibility, justifies the choice of taking into account chains obtained by repeating the same type of unit. 
In this section, we compare the asymptotic behaviour of the transport probability from the first site to the end of the chain, using \graph{C3}, \graph{C4} and \graph{DiC4} as units and with optimized phase parameters (see Fig. \ref{fig:best_chain}). In particular, we compare their performance to ascertain the improvement over the \graph{P} topology. 

\par
\begin{figure}[h!]
\centering
    \begin{subfigure}{.4\textwidth}
        \includegraphics[width =0.94\textwidth ]{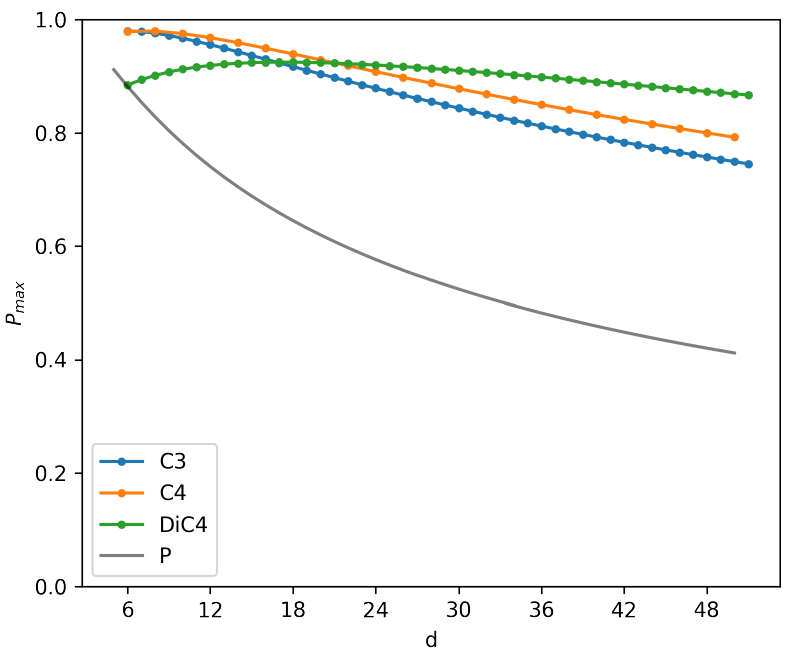}
    \end{subfigure}
    \begin{subfigure}{.4\textwidth}
        \includegraphics[width =0.95\textwidth ]{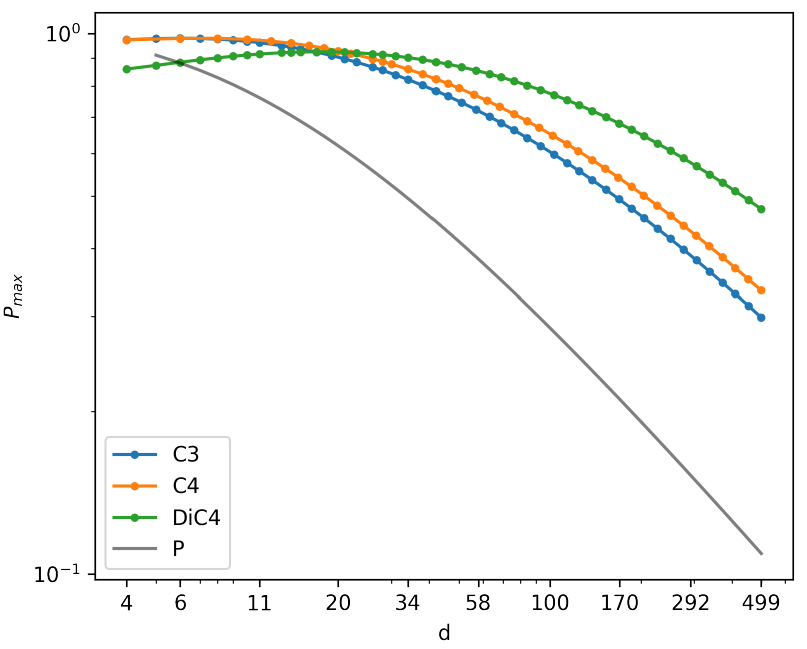}
    \end{subfigure}
\caption{
\graph{(left)} Phase optimized first maximum performance for 
optimal chains (see App.\ref{app:chains}) built from \graph{C3}, \graph{C4} and \graph{DiC4}(1,3) units, \graph{P} scaling is given as reference (black solid line).    \graph{(right)} The same plot in a wider log-log scale.
}
\label{fig:chain_scaling}

\end{figure}

 Results are illustrated in Fig.\ref{fig:chain_scaling}. The three composite chains clearly outperform the \graph{P} graphs, with a transport probability over $0.8$ at a transport distance of 50. At the same distance, the corresponding path graph does not achieve transport anymore ($p_{1 \to f} < 0.5$). Interestingly, all the curves for composite chains present an inversion of concavity meaning that, somehow counter intuitively, the best performing graphs are not the shortest ones (e.g. with $\graph{DiC4}$ units the best performing chain has $10$ units).\\

\par 
To back up these numerical observations with analytical methods, we should look more closely at $\graph{P}$ graphs and their transport performance. It is well acknowledge that \graph{P2} and \graph{P3} exhibit perfect transport between their two ends, however the transport performance of the \graph{P} class decays quickly for longer paths: their dynamics approaches that of the one-dimensional infinite lattice, whose transport probability when starting from a localized state is given by $|J_{N}(t)|^2$ at distance $N$ from the initial vertex, with $J_N$ the $N^{th}$ Bessel function of the first kind. The value of $J_N (t)$ at its first maximum $t^{*}_{1}$ has the following approximate expansion with a power law decay \cite{stegun} 
\begin{equation}
\label{eq:path_jn}
    J_{N}(t^{*}_1) \  \approx \ 0.6749 \  N^{-1/3} \ (1 -0.1617 N^{-2/3} + 0.0292 N^{-4/3}+ ...) 
\end{equation}
which gives a reference to what we might expect the asymptotic behaviour of \graph{P} graphs to be.

\par 
We can extend this analysis to our sample of chains by considering a slight variation of the \graph{P} graphs, 
which are the sped-up paths \graph{$\text{P}_{w}$} \cite{wojcik05,Banchi2011}.
Their adjacency matrices differ from those of the \graph{P} class by a uniform weight $\mu$ applied to all their edges except the first and the last one of the path.
As shown in Sec.\ref{sec:extra_kry} by going to the Krylov basis, the dynamics of our sample of chains can be shown to translate to an analogous transport problem on this class of weighted paths, where the analogy is exact for the \graph{C4} unit, while for $\graph{C3}$ and $\graph{DiC4}$ the relation between the Krylov basis and the original one becomes appreciably non-local, at least for for the last states of the Krylov basis. In addition to that, a similar mapping happens for the odd and even \graph{C} families: even \graph{C} graphs have a direct mapping to a sped-up path through Krylov transform where odd graph give inconclusive results, although showing the same kind of behaviour.  
Although we lack a formal proof, due to its similarity with the plain path graphs \graph{P} we speculate that power law expansion analogous to \ref{eq:path_jn} is valid in the whole \graph{{$\text{P}_{w}$}} family.
Expanding the modulus square in such expression one obtains a power series in the variable $N^{-2/3}$ which lacks the constant term:

\begin{equation}
    \label{eq:banchimodel}
    p_{1 \rightarrow f}(t^{*}_{1}) \approx c_1 N^{-2/3} + c_2 N^{-4/3} + ...
\end{equation}

With the coefficient $c_i$ being a function of the speedup $w$ only.
This is the formula that has been used to characterize the behaviour of our sample of transport structures.
The result of the fit are summarized in table \ref{tab:banchifit}, where we extracted the first two coefficient of the expansion \ref{eq:banchimodel}.
Notice that the fit has been obtained for graphs of a minimal length to cut off the inverted concavity behaviour at small distance, which is not directly entailed in the asymptotic model. 

\begin{table}[hbt]
    \centering
    \begin{tabular}{c||c|c|c}
         & $c_1$ & $c_2$ & $c_1/c_1^{\text{\graph{P}}}$\\
         \hline
         \graph{P} &            $7.27$ &    $-23.4$ & $1$\\
         \graph{C} even &       $1.82$ &  $-0.034$ & $0.25$\\
         \graph{C} odd &        $0.91$ &  $-0.397$ & $0.12$\\
         \graph{C3} chain &     $23.7$ &  $-304$ & $3.25$\\
         \graph{C4} chain &     $27.2$ &  $-383$ & $3.74$\\
         \graph{DiC4} chain &   $43.9$ &    $-893$ & $6.04$\\
    \end{tabular}
    \caption{Parameters extracted by fitting the phase-optimized behaviour of a selection of topologies to the spedup-line model of Eq.(\ref{eq:banchimodel}). For each family the regression has been performed on a log spaced selection of graphs with distance in the interval (500,1000)}
    \label{tab:banchifit}
\end{table}

 Although the fit might not have a direct theoretical interpretation, the coefficient of the leading term in the 
 expansion can be used to quantify the asymptotic gain in transport probability that is achieved with our chains: in the long distance limit, the ratio between the transport probabilities of two chains is given by the ratio of their $c_1$ coefficients. Our analysis shows that with the aid of non-chiral \graph{C4} chain we obtain a greater than threefold improvement in transport performance, which cannot simply be explained by the larger coupling in the Krylov basis\footnote{A larger, homogeneous coupling on a path graph leads to an obvious speedup of the whole dynamics, but not to higher first-maxima probabilities}. Chains with \graph{C3} and \graph{DiC4} units instead require a phase to attain their optimal performance: these genuinely chiral quantum walks, compared to their non chiral counterpart, offer an improvement up to six times the \graph{P} class benchmark. \\

\par
\begin{figure}[h!]
    \centering
    \includegraphics[width = .4\textwidth]{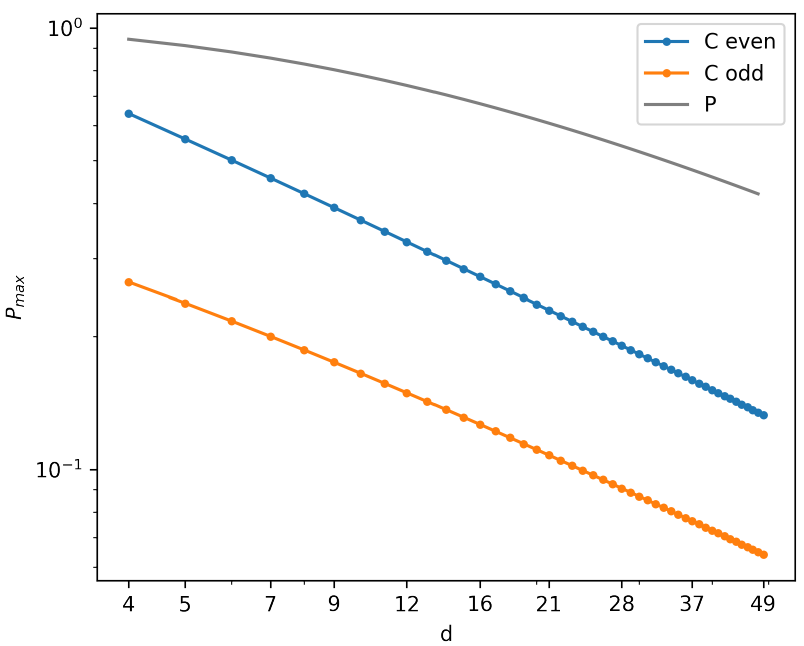}
    \caption{Phase optimized first maximum performance (loglog scale) for odd and even \graph{C} graph 
    families, \graph{P} scaling is given as reference. The only asymptotic difference between the 
    two scaling is a costant shift, which translates to a fixed ratio on linear axis.}
    \label{fig:enter-label}
\end{figure}

{Interestingly, with both single \graph{C} graphs and chains of \graph{C} units, one can identify 
two distinct behaviours between odd and even cycles (see also \cite{wojcik05}): while units with an 
even number of sites do not need the introduction of a phase to achieve their best performance, this 
is necessary in the odd case  (with total phase $\pi /2$ per unit) and all of them are thus truly
 chiral. Although there seems to be no qualitative difference in the asymptotic behaviours of the 
 transport probabilities between even and odd  units once this phase optimization has been performed 
 (see Fig. \ref{fig:enter-label}), it should be emphasized that for odd units the 
transport is favoured only in one direction, while it is inhibited in the opposed direction thanks to 
chiral effects. In principle, this could provide some enhancement at longer distances, especially if 
one considers that, as we will show in the following Sections, the mechanism by which chirality 
improves transport in chains with odd units is of a different kind with respect to changes in the 
units topology and modulations of the coupling strength by positive, real weights.}

\section{Krylov reduction of quantum transport on chain graphs}
\label{sec:extra_kry} 

\par{It is often the case that the evolution of a generic quantum walk starting from a specific state 
happens in a linear subspace, called the Krylov space, of the full Hilbert space \cite{Jafarizadeh2007,novo2015systematic}.
 With respect to the so-called Krylov basis, the evolution can be represented as a quantum walk on a 
 weighted path graph (where each edge can have a different weight), whose number of sites coincides with 
 the dimension of the subspace. Lanczos algorithm guarantees that this can be done for any 
 Hamiltonian and any starting state, always yielding a tridiagonal and real Hamiltonian in the Krylov 
 basis, corresponding indeed to a weighted quantum walk on a path graph \cite{Razzoli_2022,e23010085}. 
 
 In practice, however, this method is convenient only if the transformation between the on-site basis and the Krylov basis is local or quasi-local, meaning that each Krylov vector should have support on a small number of neighbouring sites of the original basis, and if the Krylov basis generated from the initial site is effectively of a lower dimension compared with the full Hilbert space. To study quantum transport, it is sufficient to check that the target site corresponds to the last vector of the Krylov basis (this coincidence is always true for the starting site), so that the problem of quantum transport on the original graph directly translates into quantum transport between the first and the last site of the weighted path graph, which is often easier to study. This analysis has been carried out for chains consisting of the three best performing units in their best equal phase configuration, namely \graph{C3} with $\pi /2$, \graph{C4} with no phases and \graph{DiC4(1,3)} with the $(\pi/2,-\pi/2)$ phase configuration.}\\

\par{Results for chains of composed out of \graph{C4} units with handles are simply interpreted: as the symmetry of those graph suggests, the Krylov basis consists in the "axial" nodes of the original chain and an even superposition of each couple of vertices in each square not belonging to the main chain axis.
Thus these structures ultimately behave just like path graphs, the only difference being the $\sqrt{2}$ weights on the central edges (i.e. all the edges except the first and the last one).}\\

\par
\begin{figure}[htb]

\centering
    \begin{subfigure}{.47\textwidth}
        \includegraphics[width =\textwidth ]{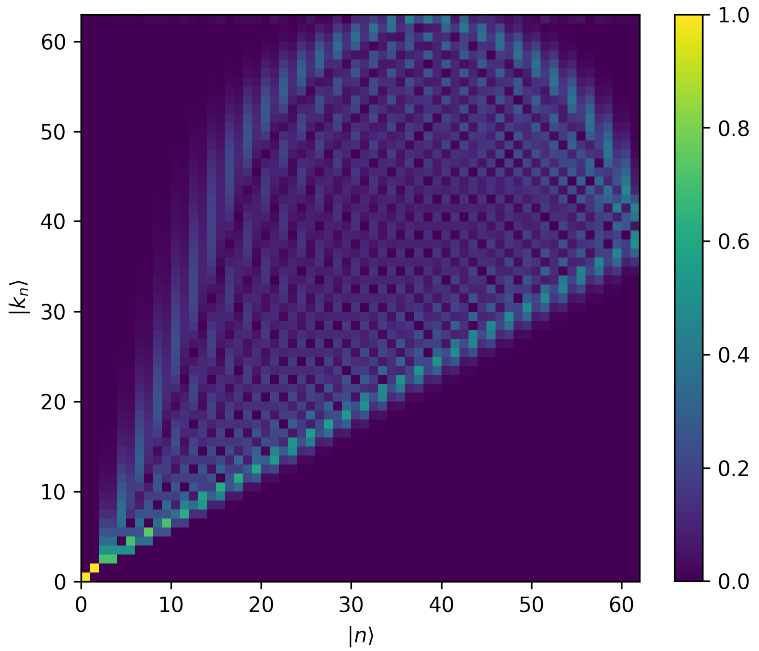}
    \end{subfigure}
    \begin{subfigure}{.50\textwidth}
        \includegraphics[width =\textwidth ]{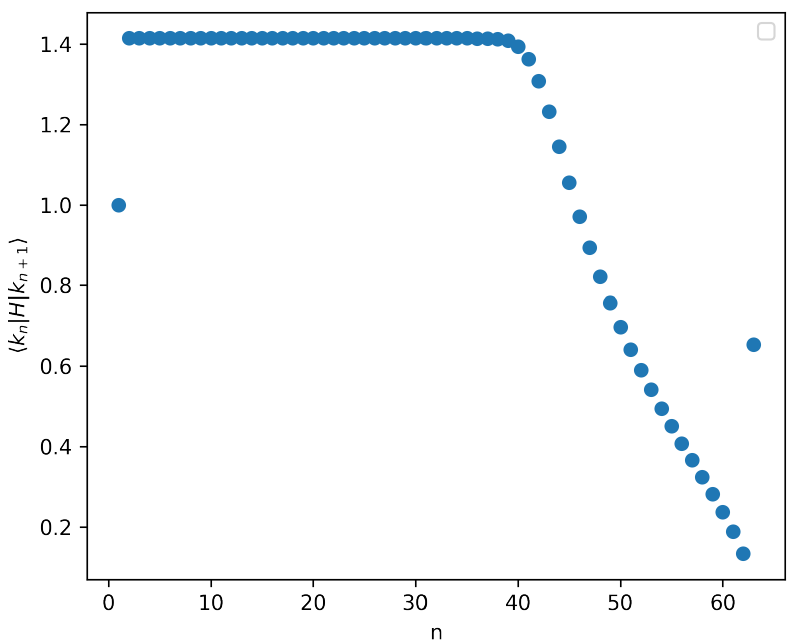}
    \end{subfigure}    
\caption{
  Modulus of overlap between Krylov basis and site basis states $\scalar{j}{k_n}$\graph{(left)} and
 couplings between the Krylov states $\bra{k_n} \mathrm{H} \ket{k_{n+1}}$\graph{(right)}  for a $\graph{C3}$ chain of 30 units with optimal phase
}
\label{fig:C3_krylov}

\end{figure}

{For chains made of \graph{C3} units, with optimal $\pi/2$ phase, Fig. \ref{fig:C3_krylov} shows that not all the Krylov vectors are sufficiently localized in the site basis. This is not surprising, 
since we are dealing with a truly chiral quantum walk, whose behavior cannot be similar to that of a weighted path graph in a local sense. However we still have a qualitative similarity, since each Krylov vector, up to the number of total units, is mostly localized around the corresponding site in the chain and we have a steady $\sqrt{2}$ speedup between the central states, which fades more gradually at the point where the Krylov states cease to be related to a clear transport event.}

Finally, a chain of kites (\graph{DiC4(1,3)} graphs) with optimal phases on each loop, i.e. two phases of $\pi/4$ in different orientation along the two triangular loops of each kite, can be reduced to a $\graph{C3}$ chain with a local change of basis with respect to the original site basis; the resulting quantum walk is still chiral, bearing the optimal phase of $\graph{C3}$, but it also has additional weights of $\sqrt{2}$ on the two edges of each triangle that don't belong to the main axis of the chain. Therefore, whereas the \graph{C4} chain is in fact equivalent to a path graph with minimal modulation of the couplings, our other two candidates, being truly chiral quantum walks on chain graphs, essentially coincide with the $\graph{C3}$ chain with optimal phases and possibly a minimal modulation of the weights.

\subsection{Krylov subspace dynamics in Cycles}

The structure of the Krylov base of \graph{C} graphs also hints at why \graph{C3}  and \graph{C4} are the best suited units to compose a chain. Again, odd and even cycles have to be considered separately due to their different behaviour. The Krylov basis for a \graph{C} graph with $2N$ sites and no phase applied is obtained straightforwardly and it consists in $N+1$ states: the initial localized state in the oroginal basis, $N-1$ even superpositions of pairs of localized states that are symmetric with respect to the starting one, and the localized state in the target site. If Lanczos algorithm is carried out with a global phase of $\pi$, instead, the Krylov basis will contain $N$ states, identical to the previous ones but missing the target site, accounting for the total suppression of transport discussed earlier in Sec.\ref{ssec:1ph_phase}.
As for the couplings in the Krylov basis, letting $\beta_{i}$ be the coupling between the Krylov state $\ket{K_i}$ and $\ket{K_{i+1}}$ one obtains:

\begin{equation}
\beta_{i} = 
\begin{cases}
  \sqrt{2}   & i = 1,N-1       \\
  1        & \mathrm{otherwise}     \\
\end{cases}
\end{equation}

\par
\begin{figure}[h!]

\centering
        \includegraphics[width =.5\textwidth ]{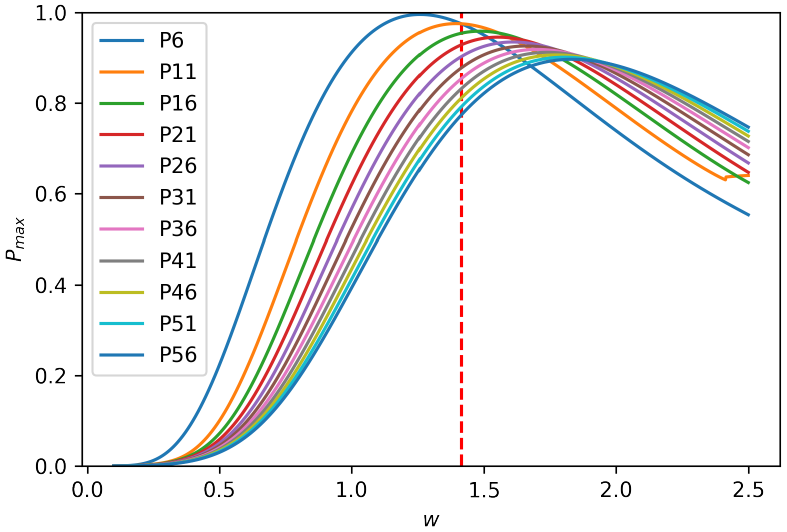}
\caption{Transport performance of first maxima as a function of speedup for a sample of \graph{$P_w$} graphs. 
The speedup parameter obtained for \graph{C3} and \graph{C4} chains ($w = \sqrt{2}$) lies around the 
optimal value of many \graph{P} graphs.}
\label{fig:optimal_coupling}
\end{figure}

{Therefore even cycles of growing number of sites behave exactly as sped-up paths $\graph{P}_{1/\sqrt{2}}$ upon rescaling the time of a factor $\sqrt{2}$. Results in Fig. \ref{fig:optimal_coupling} confirms the evidence we 
gathered through numerical analysis, i.e. that for such level of speedup the presence of a growing central part is suboptimal for transport and among this family \graph{C4} holds a special place as the only member which lacks that feature, holding a steady $\sqrt{2}$ speedup.}\\

\par{Let us now consider \graph{C} graphs with an odd number of sites and the optimal value of $\pi/2$ for the total phase. The behavior of the Krylov basis is less informative in this case: for a \graph{C(2N+1)} graph, the Krylov basis consists now of $2N+1$ states (i.e. there is no dimensional reduction), of which the first $N$ states have a spatial distribution peaked at the corresponding localized state\footnote{This means that $\vert k_{j} \rangle$ has maximal overlap with $\vert j \rangle$} and similar relative couplings close to $\sqrt{2}$. 
Those states are the ones involved in the first propagation of probability through the graph and are therefore the one relevant for the existence of the "first maxima" of probability and their relative propagation through a hypothetical chain, which is consistent with what has been observed.
One would therefore argue that the suboptimality related to a path-like dynamic still holds in the odd case, with \graph{C3} being the exceptional configuration just like \graph{C4}.} \\

\par{As the Krylov analysis has shown, most of the interesting transport behaviours found in this work are related to the transport properties on a weighted path graph, where the two extremal edge bear a weight of $1$, whereas all the other couplings are set to a certain weight $w$ (speedup parameter). These systems have been studied e.g. in \cite{Banchi2011}, where it is found that a weight scaling as $w_{opt}(N) \simeq 1.03 (N-2)^{-1/6}$ on the external edges ($N$ being the total number of vertices in the path) optimizes the asymptotic transport probability with a transport time scaling linearly with $N$ (all internal edges have unit weight). This result is displayed in Fig.\ref{fig:optimal_coupling} for our case, where the weights $w$ are on the internal edges instead and we show the transport probability at the first maximum as a function of $w$ for path graphs of various lengths. It can be seen that $w = \sqrt{2}$ provides an enhancement with respect to the unweighted path ($w = 1$), at least when the graphs are long enough. }

\section{Relation between the spectrum and quantum transport}
Seeking a better understanding of the quantum transport performances of our best candidate chiral quantum walks on chain like graphs, it is useful to follow the analysis introduced in \cite{Banchi2011}. 
Whenever the Hamiltonian $\mathrm{H}$ is real, it can be diagonalized by an orthogonal matrix $\mathrm{O}$, such that $\mathrm{O}^{T} \mathrm{H} \mathrm{O} = \mathbf{\Lambda}$ and $\mathbf{\Lambda} = \mathrm{diag} ( \lambda_{1} , ..., \lambda_{N})$. Moreover, for chain-like graphs, the mirror symmetry that reflects the chain around a plane perpendicular to it and passing through the middle vertex, there is a relation of the type $\mathrm{O}_{k,j} = \pm \mathrm{O}_{k,N+1-j}$ for $j = 1, ..., N$, $N$ being the total number of vertices; if $n$ labels the eigenvalues in increasing order, the relative sign is alternating with $n$; therefore, the transport probability from the first to the last site becomes:
\begin{equation}
\label{eq:ptranspbanchi}
P_{1 \to N} (t) \ = \ \vert \sum_{n} R_{n} e^{i \pi n -i \lambda_{n} t} \vert^{2}
\end{equation}
where $R_{n} = \vert \mathrm{O}_{n,1} \vert^{2}$ and therefore $\sum_{n} R_{n} = 1$. In order for this to be close to $1$, the phases should all be approximately equal at a certain time: this implies that $\lambda_{n} = \nu n + s_{0}$ for some constant shift $s_{0}$ and some speed $\nu$. It is therefore clear that whenever the chain has a mirror symmetry and the Hamiltonian is real, optimal transport from the first to the last site requires an equispaced spectrum, of the type that can be achieved with appropriate weights as in the Krawtchouk chain \cite{Groenland18}. In general, minimal modulation of the weights (e.g. the weighted path discussed before) has the effect of modulating the values of $R_{n}$ around the centre of the spectrum, where $\lambda_{n} = 0$, whereas the spectrum itself is essentially unaffected\footnote{When $N$ is large, only two edges out of $N$ have different weights and the spectrum will be close to that of the unweighted path; on the other hand, the coefficients $R_{n}$ are strongly influenced by the matrix elements of $\mathrm{O}$ involving the first and the last sites, and therefore will significantly change with the modulation of the weights on the extremal edges.} for large $N$ and it resembles the spectrum for the unweighted path; therefore, by appropriately choosing the weight $w$, one can tweak the values of $R_{n}$ so that only those integers $n$ for which $\lambda_{n}$ is linear with $n$ give a significant contribution to the sum in Eq.(\ref{eq:ptranspbanchi}), which was  the main idea exploited in \cite{Banchi2011}.  \\ 

\begin{figure}[!ht]
\begin{center} \includegraphics[width=0.4\textwidth ]{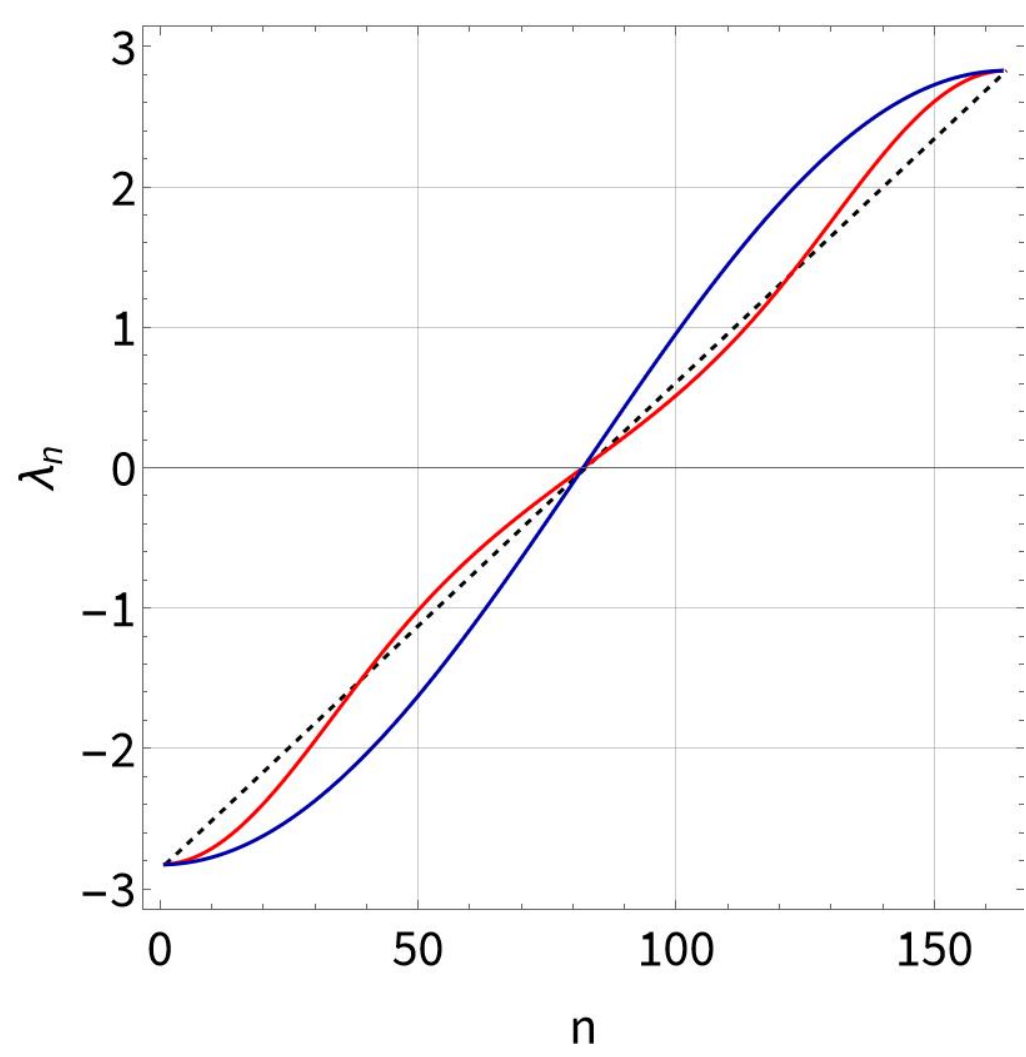}
\end{center}

\caption{Spectra of Hamiltonians for path graph (dark blue) with 163 sites and for the \graph{C3} chain with handles and optimal phase with $80$ triangle units (red). The total number of vertices is the same for both graphs. $n$ labels the eigenvalues sorted in increasing order. The dotted black line shows the linear trend for reference. Notice that minimal modulation of weights on the path graph will not visibly change the blue curve. }
\label{fig:spectrachains}

\end{figure}

Figure \ref{fig:spectrachains} shows that the \graph{C3} chain with handles and with the optimal phase has a spectrum which is closer to a linear one, when compared with the spectrum of \graph{P} or its variation with modified, equal weights on the first and last edge: interestingly, then, although minimal modulation of the weights cannot significantly alter the spectrum in favor of transport for large $N$, chiral effects do help  in this regard, without breaking the homogeneity of the chain. However, the \graph{C3} optimal chain is chiral: its Hamiltonian is not real and some of its eigenvectors are necessarily complex. For these reasons, looking at the linearity of the spectrum is not fully sufficient to deduce a transport enhancement in itself.

In general, letting $\mathrm{U}_{\mathrm{H}}$ be the unitary matrix that diagonalizes the Hamiltonian $\mathrm{H}$ for the chiral triangle chain and $\lambda_{k}$ its eigenvalues, the transport amplitude will be:
\begin{equation}
A_{1 \to M} (t) \ = \ \sum_{k=1}^{M} \mathrm{U}^{*}_{k,M} \mathrm{U}_{k,1} e^{-i \lambda_{k} t} 
\end{equation}
where $N$ is the number of triangle units and $M = 2N+3$ is the total number of sites (therefore site $2N+3$ is the target for transport).
Let us call $\phi_{k} = \mathrm{arg} ( \mathrm{U}^{*}_{k,M} \mathrm{U}_{k,1} )$. We see that the condition of optimal transport at time $\tilde{t}$ requires:
\begin{equation}
\label{eq:condphasetransp}
 \lambda_{k} \tilde{t}  \ \simeq \ \phi_{k} + s_{0}  \ \ \ \ \ \forall k = 1,...,M
\end{equation}
where $s_{0}$ is a possible constant phase shift.  \\

For the optimal value of the total phase on the \graph{C3} units, the Hamiltonian of the chain has a characteristic symmetry that allows us to better explore the condition in Eq.(\ref{eq:condphasetransp}). Indeed one can easily check that a reflection about the middle of the chain (defined by a permutation matrix $\Pi$) combined with an appropriate quasi-gauge transformation $\mathbf{U}_{D}$ gives a unitary transformation $\mathrm{U}_{S} = \mathrm{U}_{D} \Pi $ such that:
\begin{equation}
\label{eq:symmH}
\mathrm{U}_{S}^{\dagger} \ \mathrm{H} \  \mathrm{U}_{S} \ = \ - \mathrm{H}
\end{equation}
This equation implies, in particular, that the spectrum of $\mathrm{H}$ is symmetric around the zero of the energy. Clearly:

\begin{equation}
A_{1 \to M} (t) \ = \ \langle M \vert e^{- i \mathrm{H} t} \vert 1 \rangle \ = \ \langle M \vert \mathrm{U}_{S}^{\dagger} e^{i \mathrm{H} t} \mathrm{U}_{S} \vert 1 \rangle \ = \ e^{i \varphi } \langle 1 \vert e^{i \mathrm{H} t } \vert M \rangle \ = \ e^{i \varphi } A_{M \to 1} (-t)
\end{equation}
where $\varphi$ is a constant phase determined by $\arg(\mathrm{U}_{D})_{1,1} - \arg(\mathrm{U}_{D})_{M,M}$ and we exploited the fact that $\Pi \vert 1 \rangle = \vert M \rangle$ and $\Pi \vert M \rangle = \vert 1 \rangle$. But now notice that $A_{M \to 1}(-t)$ is the complex conjugate of $A_{1 \to M} (t)$, therefore the previous chain of equalities leads us to conclude that:
\begin{equation}
A_{1 \to M} (t) \ = \ e^{i \varphi / 2} \sqrt{P_{1 \to M} (t) }
\end{equation}
or in words, that the phase of the transport amplitude is constant in time. It is important to stress that this is a consequence of the special symmetry of $\mathrm{H}$ in Eq.(\ref{eq:symmH}): whenever the value of the phase is different from the optimal one, this symmetry will not be respected. Notice that if there exists a time $\tilde{t}$ such that $\forall k: \phi_{k} - \lambda_{k} \tilde{t} = s_{0}$, then $s_{0} = \varphi /2$ would be the constant phase of the amplitude; since we can argue that this constant phase must be zero, we conclude that we can assume $s_{0} = 0$ in Eq.(\ref{eq:condphasetransp}).  
The same reasoning can be applied to the transport amplitudes $A_{k \to M- k} (t)$ for a walker initially localized at site $\vert k \rangle$ to be found in the localized state $\vert M - k \rangle$ at time $t$: also in this case the phase will be constant in time. Therefore, provided that the transport probabilities are optimized, these chains also allow for transport of superposition states of a few sites of the chain. As was recently argued \cite{entanglementrouting} for a spin-chain analogue model, the transport of superposition states can be exploited to transport entanglement; indeed, our argument about the time-independence of the amplitude's phase also explains the results in \cite{entanglementrouting}, at the same time backing up the role of chirality in achieving this property. On the other hand, if the goal is the transport of a quantum bit, the pure quantum walk model encoding the qubit state in a spatial superposition of the first two sites of the chain-graph can be more convenient in a practical implementation, since it requires a much smaller Hilbert space, thus being less sensitive to decoherence sources \footnote{Indeed, in the spin-chain model, coherent transport would require to preserve both the spatial coherence between different sites \emph{and} the local coherence in the spin basis.  }.


\section{Conclusions}

\par{In this paper, we have addressed the enhancement of quantum transport on chain-graphs through the use of chiral quantum walks. We have studied how the topology of graph units, and the consequent phases configurations, affects the probability of transport and the speed of the process, establishing convenient criteria to define transport events 
and asses the performance of a diverse set of structures on a common ground. In particular, we have determined that 
the first maximum of the transport probability in time is always a transport event (i.e. with probability higher 
than the threshold of $1/2$) when transport is possible and the phase configuration are optimized. Looking for higher maxima in a time window leads to no substantial improvement in the short term, while leading to a chaotic behaviour when the time window gets too wide. Moreover, these \emph{first maxima} events display a general linear relation between transport time and topological distance, with a slope tightly linked to the topology.}\\

\par{The effect of free phases on transport have been specifically addressed in Sec.\ref{ssec:1ph_phase}. For one-loop graphs, a complete characterization of their phase response had been possible: cycles behave best with a specific phase value, which has been shown to be always $0$ if the number of sites is even, and $\pi/2$ otherwise. This phase criterion has been extensively tested numerically, and it was shown to hold also for all kind of chain-like combination of those simple loops. In all those cases, the choice of the optimal phases for each unit of the chain always results in a relevant transport event, always coinciding with the first maximum.}
Among all the considered single graph units, three stood up for having the fastest transport time and at the same time the best performance in terms of transport probability when combined in chains of arbitrary length: these are \graph{C3}, \graph{C4} and \graph{DiC4} units, with $\graph{C4}$ being the only one not requiring a phase to achieve its best performance. 

To complement the numerical results, we have employed Krylov reduction to show that the transport behaviour of \graph{C} and \graph{h(C)} graphs are ultimately reducible to the one observed for the \graph{P} graphs with minimal modulation of the weights on the edges \cite{Banchi2011}. 
On the other hand, we could not argue the same for the chiral $\graph{C3}$ and $\graph{DiC4}$ units, whose transport peculiarities we studied by examining their spectra and eigenvectors support, concluding that their advantage stems from an inherently chiral effect, of a different kind than minimal modulation of weights on path graphs. 
Although looking at Fig.\ref{fig:chain_scaling} we cannot claim that these classes of chain offer a completely 
new kind of transport scaling, the performance of first maxima offers a considerable improvement over the \graph{P} 
class reference, leading to transport of the initial state with a probability above $0.8$ even at a topological distance of around $50$. As a final remark, we have shown that our chiral chains enjoy a special symmetry entailing a phase stability of the transport amplitude, thus making them amenable to the transport of quantum superposition states \cite{10.1116/5.0146805}.  

\par{Our results widen the scope of research in quantum transport, suggesting to harness chiral effects and providing promising results on their role. Some further research would be necessary to determine if a more tackling numerical optimization over the whole space of phases parameters of a chain graph can provide any further advantage. 
Additionally, it would be interesting to clarify whether non-classicality of evolution \cite{PhysRevA.102.012201} 
is somehow linked to transport performance, and if modulation of the weights can be efficiently combined with 
chiral effects to achieve even better performances.}

\appendix

\section{Chain structures}
\label{app:chains}

\begin{figure}[h!]
    \centering
 \includegraphics[ width =\linewidth ]{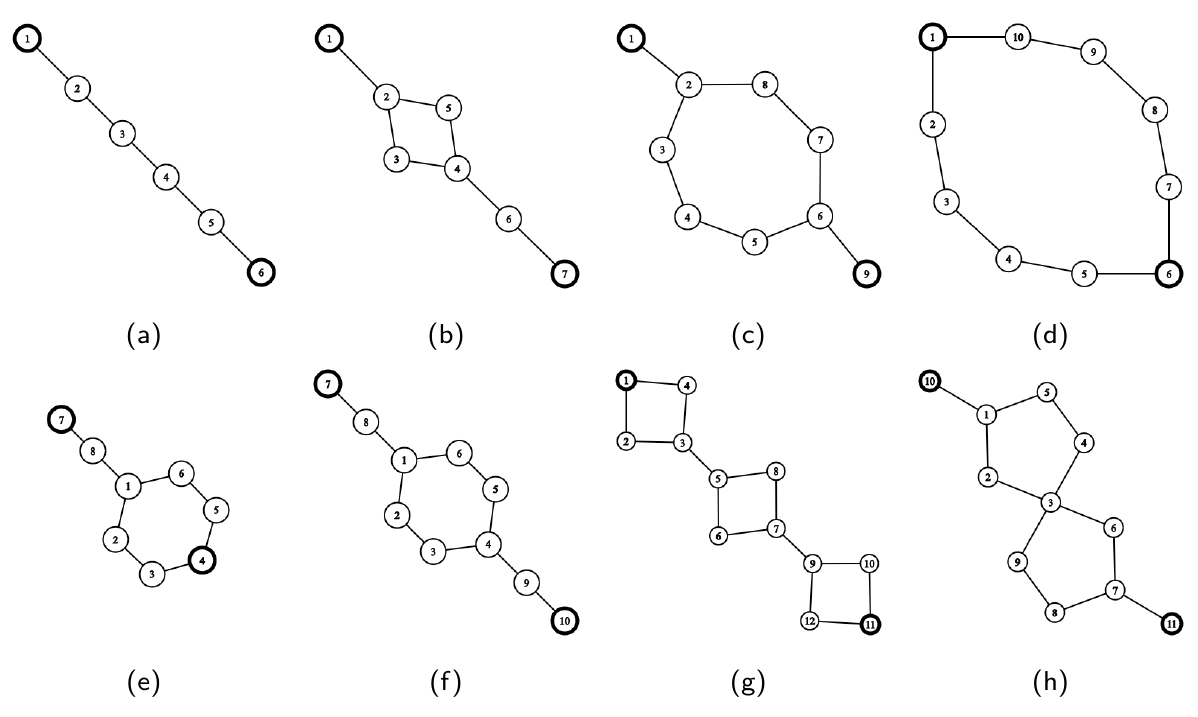}    
    \caption{ Representative set of the various configuration tested for graph composition, start and target site have been highlighted.
    In order: a) \graph{L6}, b) \graph{L2/C4/C3}, c) \graph{h(C7)}, d) \graph{C10}, e) \graph{L3/C6}, f) \graph{L3/C6/L3}, g) \graph{C4+C4+C4}, e) \graph{h(C5/C5)} }
    \label{fig:chain_example}
\end{figure}

The results gathered in this work suggest that there are 2 ways to improve the transport performance of 
a Chiral Quantum Walk, and that they work independently. The first one is the right use of phase parameter 
of a given graph which has already been discussed, the second one is a wise choice of the structure 
considered for transport. Although our set of graphs allows for many different way to assemble basic graphs into longer chains,  extensive numerical trials and a systematic analysis of a wide array of features, hinted at the 
existence of an optimal criterion that can guide graph creation.
Following the idea that the introduction of loops produce different and possibly better transport dynamic, 
the features that have been analyzed concern (see Fig. \ref{fig:chain_example}):
\begin{itemize}
    \item {Given a transport distance, which fraction of the path should be substituted with a single loop to 
    obtain the best performance (\graph{a} to \graph{d})} 
    \item {Given a transport distance, which fraction of the path should be substituted with multiple smaller 
    loops to obtain the best performance (\graph{g},\graph{h})}
    \item {Chosen central unit, which handles configuration (attaching \graph{P} graphs at transport ends) 
    works best(\graph{e},\graph{f} or \graph{c})}
    \item {Given multiple loop unit on a chain, what is the best edge distance to concatenate them 
    (\graph{g}, \graph{h})}
\end{itemize}

Our results may be summarized in the following criterion: state transfer at great distance is best tackled by 
small unit from the pool of basic graph (e.g. \graph{C3} or \graph{DiC4}) joined in a chain using the operation 
"/", that is without adding any extra edge. The addition of handles ( " $h(\cdot)$ " operator) makes those 
kind of structure optimal. This criterion underlies the choice of the three best chain structures discussed 
in this work (see Fig. \ref{fig:best_chain}) and has been referred to as {\em chains} throughout the paper. 
A\graph{C3} chain of 5 units therefore corresponds to \graph{h(C3/C3/C3/C3/C3)}. Any particular chain 
is thus addressed just by the unit chosen and the number of repetition: no new notation has been developed 
for this specific operation and the one used for simpler graphs, though expressive, gets quickly cumbersome.

\acknowledgements 
This work has been partially supported by MUR and EU through the project PRIN-2022-PNRR QWEST (P202222WBL). 
MGAP is member of INdAM-GNFM. The authors thank Alberto Bottarelli, Claudia Benedetti, Alessandro Candeloro, 
Luca Razzoli, Simone Cavazzoni, Paolo Bordone, and Stefano Olivares for discussions.

\bibliography{chains}
\bibliographystyle{ieeetr}

\end{document}